\documentclass[american,aps,prb,twocolumn,amsmath,amssymb,showpacs,
superscriptaddress,notitlepage,longbibliography]{revtex4-2}
\usepackage{float}

\usepackage[T1]{fontenc}
\usepackage[utf8]{inputenc}

\usepackage{babel}

\usepackage{xcolor}
\usepackage{booktabs}
\usepackage{mathrsfs}
\usepackage{amsmath}
\usepackage{amssymb}
\usepackage{stmaryrd}
\usepackage{graphicx}
\usepackage{esint}
\usepackage{float}

\newcommand{\Masoud}[1]{{\color{black}#1}}
\makeatletter
\newcommand{\lyxmathsym}[1]{\ifmmode\begingroup\def\b@ld{bold}%
  \text{\ifx\math@version\b@ld\bfseries\fi#1}\endgroup\else#1\fi}
\providecommand{\tabularnewline}{\\}
\makeatother

\usepackage{bbm}
\PassOptionsToPackage{pdfencoding=auto,psdextra,colorlinks=true,linkcolor=blue,anchorcolor=red,citecolor=blue,urlcolor=blue}{hyperref}
\usepackage{hyperref}
\usepackage{xurl}
\hypersetup{breaklinks=true}

\usepackage[caption=false]{subfig}
\usepackage{lipsum}
\usepackage{balance}

\allowdisplaybreaks

\AtBeginDocument{%
}

\begin{document}

\title{Beyond spin-1/2: Multipolar spin-orbit coupling\\ in noncentrosymmetric crystals with time-reversal symmetry}
\author{Masoud Bahari}
\email{masoud.bahari@uni-wuerzburg.de}
\affiliation{Institute for Theoretical Physics and Astrophysics$\text{,}$ University
of Würzburg, D-97074 Würzburg, Germany}
\affiliation{Würzburg-Dresden Cluster of Excellence ctd.qmat, Germany}
\author{Kristian M{\ae}land}
\affiliation{Institute for Theoretical Physics and Astrophysics$\text{,}$ University
of Würzburg, D-97074 Würzburg, Germany}
\affiliation{Würzburg-Dresden Cluster of Excellence ctd.qmat, Germany}
\author{Carsten Timm}
\affiliation{Würzburg-Dresden Cluster of Excellence ctd.qmat, Germany}
\affiliation{Institute of Theoretical Physics, Technische Universität Dresden,
01062 Dresden, Germany}
\author{Björn Trauzettel}
\affiliation{Institute for Theoretical Physics and Astrophysics$\text{,}$ University
of Würzburg, D-97074 Würzburg, Germany}
\affiliation{Würzburg-Dresden Cluster of Excellence ctd.qmat, Germany}
\date{\today}

\begin{abstract}
We develop a symmetry-adapted multipolar $\mathbf{k}\cdot\mathbf{p}$ theory close to the bulk $\Gamma$ point for time-reversal-symmetric, noncentrosymmetric $C_{3v}$ crystals in the strong atomic spin-orbit-coupling ($jj$-coupling) limit. Using a $j\in\{1/2,3/2,5/2\}$
multiplet basis appropriate for heavy-element \textit{p}- and \textit{d}-bands, we systematically construct all symmetry-allowed spin-orbit coupling terms up to fifth order in momentum and generalize the usual spin texture to a total-angular-momentum texture. For $j>1/2$, multipolar spin-orbit coupling qualitatively reshapes Fermi surfaces and makes the topology of Bloch states band dependent. This leads to anisotropic high-$j$ textures that go beyond a single Rashba helix.
We classify these textures by their total-angular-momentum vorticity
$W_{n}$ for every energy band and identify distinct $|W_{n}|=1,2,5$ phases.
We show that their crossovers generate enhanced and nonmonotonic current-induced
spin-polarization responses, namely the Edelstein effect, upon tuning the chemical potential. Our results provide a symmetry-based framework for analyzing and predicting multipolar spin-orbit coupling,
total-angular-momentum textures, and spintronic responses in heavy-element materials without an inversion center.
\end{abstract}

\maketitle

\section{Introduction}

Spin-orbit coupling (SOC) in crystals without inversion symmetry gives
rise to spin-split bands and momentum-dependent spin textures that
underpin a broad range of phenomena in contemporary condensed-matter
physics \citep{RevModPhys_1982,RevModPhys_1,Barnes2014,RevModPhys_2017,RevModPhys_2021,RevModPhys_2022,RevModPhys_2023,RevModPhy_2024,Texture_2008,Spin_dynamic_2010,Texture_2010,Texture_2012_2,Texture_2015_2,Texture_2015_5,Texture_2017_5,Texture_2019_5,RevModPhys_1,Blugel_2017,RALPH_2008,RevModPhys_2019,2006ApPhL_2006,Yoda2015}. This ranges from topological insulators and semimetals to spintronic
interfaces and unconventional superconductors \citep{Texture_2008,Spin_dynamic_2010,Texture_2010,Texture_2012_2,Texture_2015_2,Texture_2015_5,Texture_2017_5,Texture_2019_5,RevModPhys_1}.
In such systems, the expectation value of the spin as a function
of crystal momentum defines a texture on the Fermi surface that directly
controls magnetoelectric responses \citep{Blugel_2017}, spin torques
\citep{RALPH_2008,RevModPhys_2019}, and the conversion between charge
and spin currents~\citep{2006ApPhL_2006,Yoda2015}. 

On surfaces and in bulk materials with $C_{3v}$ symmetry, SOC effects
are particularly rich. Beyond the paradigmatic linear Rashba model,
higher-order terms in momentum generate hexagonally warped Fermi surfaces
and spin textures with sizeable out-of-plane components, as extensively
discussed for topological-insulator surfaces and Rashba alloys such
as $\text{Bi\ensuremath{_{2}}Te\ensuremath{_{3}}}$ and Bi/Ag(111)
\citep{SOC_2009b,SOC_2012b,SOC_2014,SOC_2018b}. Recent work has shown
that these higher-harmonic SOC contributions and the associated winding
patterns of the spin texture are generic in systems with strong SOC
and broken inversion symmetry, appearing in both surfaces and bulk
crystals, including ferroelectric oxides and van der Waals heterostructures
with radial Rashba fields \citep{Texture_2018_2,Texture_2020_4,Texture_2022_1,Texture_2024_4,Texture_2024_5}.
This motivates a systematic treatment of SOC beyond the Rashba
form, especially in three-dimensional materials, where the full point-group
symmetry and the $k_{z}$ dependence of the dispersion play an essential
role.

Most theoretical analyses of spin-orbit-coupled materials, however,
still build on single-band or effective spin-$1/2$ models.
In this case, the orbital character is weak and the internal (spin) degree
of freedom is encoded in the spin texture $\langle\boldsymbol{\sigma}\rangle$
of the Bloch bands \citep{Texture_2012,Texture_2016_5,Texture_2017_2,SOC_2018b,Texture_2018_2,Texture_2020_3,Texture_2020_7,Texture_2020_5,Texture_2022_1,Texture_2024_4}.
This picture implicitly treats the spin as an isolated two-level system
and works well as long as orbital degrees of freedom and higher total
angular momentum (TAM) components can be ignored. However, in many
materials involving heavy \textit{p}-, \textit{d}-, or \textit{f}-electron elements where SOC is strong, this assumption breaks down. Then, spin is no longer a good quantum number. Nevertheless, the TAM is typically a good quantum number in these materials.
Consequently, the low-energy bands can be associated with their TAM $j$, which can be larger than $1/2$, e.g., $j=3/2$ or $j=5/2$ \citep{GroupThe_1955,Luttinger1956,Winkler2003,Circ_2004}. 

While many nonmagnetic semiconductors are well described by effective spin-$1/2$ bands, materials built from lighter elements can also realize $j=3/2$ low-energy manifolds. This occurs, for instance, near fourfold band degeneracies in cubic crystals, whereas higher-multiplicity crossings are comparatively rare. Such higher-fold degeneracies can nevertheless be enforced by symmetry in certain nonsymmorphic space groups~\citep{Beyond2016}. 

In the spin-$1/2$ case, the Pauli matrices satisfy $\hat{\sigma}_{i}^{2}=1$ and the product of two different Pauli matrices is the third one. In this case, the set of the identity matrix and the Pauli matrices is closed under multiplication and prohibits building non-trivial higher-multipolar operators. By contrast, in higher-$j$ manifolds, the angular-momentum operators obey $\hat{J}_{i}^{2}\neq1$ and the products of angular-momentum matrices are higher multipoles. Therefore, the spin density matrix acquires a genuine multipolar structure containing quadrupolar, octupolar, and higher-rank components \citep{Brydon2016,Texture_2017_4,Kim2018,Boettcher2018,Roy2019,Menke2019,Sim2020,Texture_2021_4,Texture_2024_6}.

For $j>1/2$ electrons, the natural analogue of the familiar spin texture
is the texture of the TAM, defined by the expectation value $\langle\hat{\mathbf{J}}\rangle$
in the Bloch eigenstates. It characterizes how the average TAM vector
(or, more generally, the corresponding set of multipoles) varies as
a function of crystal momentum. Clarifying the distinction between conventional spin$-1/2$
textures $\langle\hat{\boldsymbol{\sigma}}\rangle$ and TAM textures
$\langle\hat{\mathbf{J}}\rangle$ is therefore crucial for correctly
capturing the momentum-space structure of internal degrees of freedom.

In this work, we develop a general multipolar SOC description in time-reversal-symmetric,
non-centrosymmetric $C_{3v}$ crystals, where we treat the TAM explicitly
in the multiplets $j \in \{1/2,3/2,5/2\}$. Using point group symmetry together with time-reversal symmetry (the gray group $C_{3v}\!+\!T$) and a double-group representation analysis,
we systematically construct all SOC terms allowed near the $\Gamma$
point, up to fifth order in momentum, by combining momentum polynomials
and TAM-tensor matrices that transform as the trivial irreducible
representation (irrep). This procedure yields a modified Rashba channel that
is present for all $j$, together with genuinely high-rank multipolar
SOC terms that exist only for $j>1/2$. The latter acts within and
between different $j$ sectors and introduce $m_{j}$-dependent energy
shifts and multipolar analogues of Rashba coupling that cannot be
captured by effective spin-$1/2$ models. 

We demonstrate that these multipolar
SOC terms naturally generate Fermi surfaces with nontrivial spin textures
that go far beyond a single helical Rashba pattern. In particular,
the interplay between linear, cubic, and quintic SOC invariants gives
rise to vorticity phase diagrams in which the TAM texture exhibits
conventional single winding, as well as double and fivefold windings
around the $\Gamma$ point, depending on the momentum and the relative
strength of the SOC channels. 

Furthermore, we show that the higher-rank multipolar couplings
can split heavy- and light-mass bands, induce strong interband hybridization,
and give rise to highly anisotropic TAM textures with unconventional
winding numbers (e.g., $W\in\{2,5\}$). In the corresponding
TAM-vorticity phase diagram, the heavy-mass bands dominated by $|j,m_{j}\rangle$
states with $|m_{j}|>1/2$ are markedly more anisotropic
than the light-mass $|j,\pm1/2\rangle$ bands. This enhanced
anisotropy arises from the emergence of multipolar SOC terms, which
generate higher-order $\mathbf{k}$ dependence of the effective spin-orbit
field.  

Moreover, within a semiclassical Boltzmann framework, we then compute the Edelstein
susceptibility for the multiband textures generated by our $\mathbf{k}\cdot\mathbf{p}$ model \citep{Edel_1990,Edel_2001,Edel_2011,Edel_2014}. We demonstrate
that high-$j$ multiplets and multipolar SOC lead to a strongly enhanced
and non-monotonic charge-to-TAM response as a function of chemical
potential. This contrasts with the smooth behavior of the conventional Rashba model. In particular, we find plateau-like structures and moderate
changes in the Edelstein tensor whenever the Fermi level crosses between
different $j$ multiplets.

Particularly appealing platforms in this context are non-centrosymmetric
compounds such as $\text{PtBi}_{2}$ and $\text{BiTeI}$, both realizing
bulk $C_{3v}$ point group symmetry \citep{Ishizaka2011,PtBi2_2019,PtBi2_2021,PtBi2_2023,PtBi2_2024,Multipolar_signature,Texture_2024_8,SOC_2025c,Changdar2025}.
The former (latter) exhibits Weyl semimetal and unconventional superconductivity
phases \citep{Kuibarov2024,Changdar2025,Xiaochun2025,Maeland_2025, SOC_2025b, Maeland2025Dec}  (giant bulk Rashba-type SOC \citep{Ishizaka2011}).
For $\text{PtBi}_{2}$, density-functional theory reveals strongly
spin-orbit-coupled \textit{p}- and \textit{d}-orbital-derived multiplets in the vicinity
of the Fermi energy, which are naturally organized into $j=3/2$
and $j=5/2$ manifolds \citep{Texture_2024_8}. In $\text{BiTeI}$,
by contrast, the low-energy bands are dominated by Bi and Te/I $p$
states that form the canonical $j=1/2$ and $j=3/2$
multiplets of a giant Rashba semiconductor. Taken together, these
materials are candidates for realizing multipolar and complex three-dimensional
TAM textures. 

Our analysis identifies the distinct roles of
$m_{j}$-dependent energy shifts and multipolar Rashba terms in shaping
the Edelstein response and suggests that $C_{3v}$ materials with
strong SOC, such as $\text{PtBi}_{2}$, are promising platforms for
engineering large and tunable Edelstein effects, complementing existing
proposals based on interfaces, Weyl semimetals, and two-dimensional
materials \citep{Edel_2018,Edel_2019,Edel_2020_2,Texture_2020_4,Texture_2024_4,Edel_2025_1,Edel_Ptbi}.

The remainder of this paper is organized as follows. In Sec.~$\text{\ref{sec:Model-Hamiltonian}}$,
we develop the symmetry-constrained $\mathbf{k}\cdot\mathbf{p}$ Hamiltonian for time-reversal-symmetric,
non-centrosymmetric $C_{3v}$ systems, working in the $j\in\{1/2,3/2,5/2\}$
basis. Symmetry properties of our model are discussed in Sec.~$\text{\ref{sec:Symmetry-properties}}$.
In addition, in Sec.~$\text{\ref{sec:Band-Basis-representation}}$,
we propose a comprehensive analysis of the spin-split spectrum in $j>1/2$ basis. In Sec.~$\text{\ref{sec:Winding-of-total}}$,
we analyze the resulting Fermi-surface TAM textures and vorticity
phase diagrams, highlighting the emergence of higher-winding TAM textures
and their manifestations in warped constant-energy contours. In Sec.~$\text{\ref{sec:Current-induced-spin-polarizatio}}$, 
we compute the Edelstein tensor within a semiclassical Boltzmann approach
and show how multipolar SOC enhances and reshapes the current-induced
TAM polarization as the chemical potential is tuned. Finally, in Sec.~$\text{\ref{sec:Conclusion}}$,
we summarize our main results and discuss experimental implications
for related $C_{3v}$ materials. The decomposition of $j$ into the appropriate irreps is given in Appendix~\ref{Appnd3}.  Further details on the transformation of the high-$j$ basis and momentum under the point group are provided in Appendix~\ref{Appnd1}. The complete SOC model is presented in Appendix~\ref{Appnd2}. A summary of effective model Hamiltonians for $j=1/2$ and $j\in\{ 3/2,5/2\}$ are given in Appendix~\ref{All_Hamils}. In Appendix~\ref{FPCs}, we explain how one can compute TAM texture from first-principle calculations.


\section{\label{sec:Model-Hamiltonian}Model Hamiltonian}
 
We analytically derive a $\mathbf{k}\cdot\mathbf{p}$ model Hamiltonian
to quantify the strong SOC near the $\Gamma$ point in the three-dimensional (3D) bulk Brillouin
zone of a system with $C_{3v}$ point group symmetry and time-reversal
symmetry. The system features three mirror planes $\{\sigma_{v_{1}},\sigma_{v_{2}},\sigma_{v_{3}}\}$
intersecting at a threefold rotational axis $\{C_{3z},C_{3z}^{2}\}$,
forming a ditrigonal pyramidal structure with $C_{3v}$ point group
symmetry \citep{GroupThe_1963,GroupThe_1988}. In the strong atomic spin-orbit regime, we employ $jj$ coupling, where $l$ and $s$ are entangled into $j=l+s$ and the $l$ manifold splits into $j\in\{|l-1/2|,l+1/2\}$. We then project onto the $j=3/2$ (\textit{p}-derived) and $j=5/2$ (\textit{d}-derived) subspaces relevant near the Fermi level.

\begin{figure}[t]
\centering{}\includegraphics[scale=0.47]{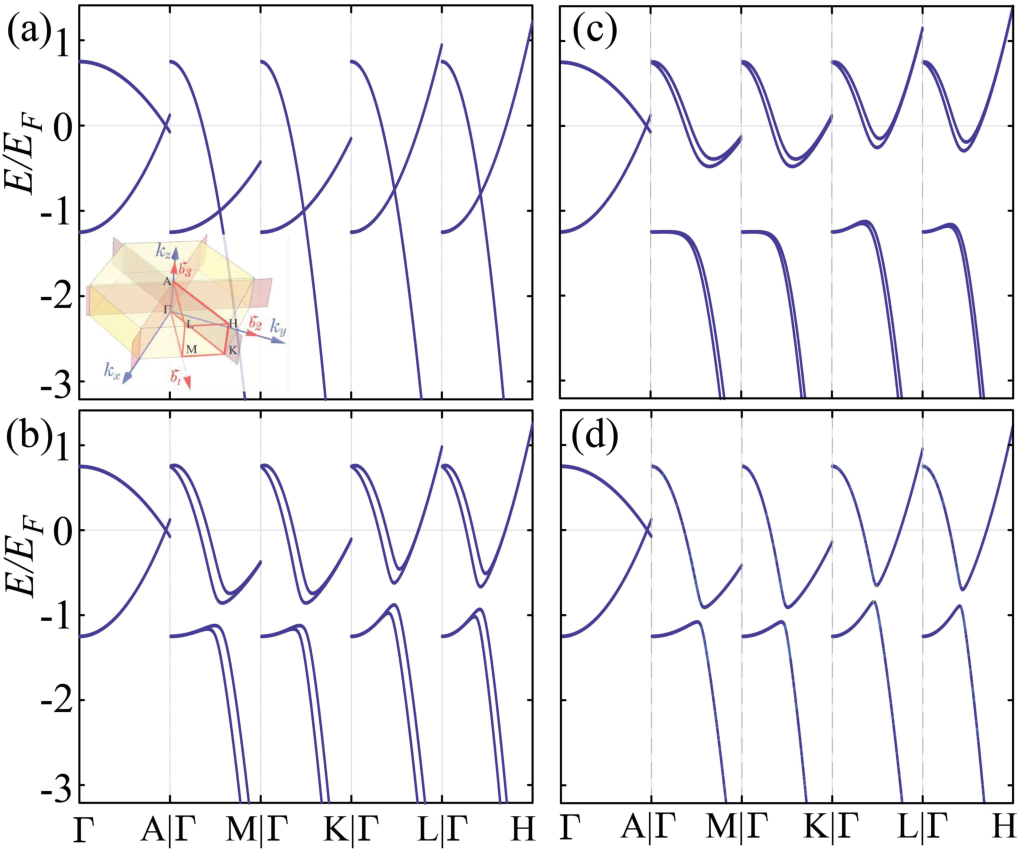}
\caption{\label{fig:Bulk-band-dispersion}Bulk band dispersion for $j=3/2$
electrons near $\Gamma$ point preserving time-reversal and $C_{3v}$
symmetry, shown for two high-rank spin-orbit coupling configurations from decomposition (a) $A_{1,+}\otimes A_{1,+}$ and (b)--(d) $A_{1,+}\otimes A_{1,+}+E_{-}\otimes E_{-}$. Model parameters are (a)  $(\mu_{2},\alpha_{2,1},\alpha_{2,2})/E_{F}=(-1,1,1)$, (b) $\gamma_{1,1}=0.3/E_{F}$, (c) $\gamma_{1,2}=0.6/E_{F}$,
and (d) $\gamma_{1,3}=0.3/E_{F}$. Other parameters are $(\alpha_{1,1},\alpha_{1,2})/E_{F}=(-2,-1)$ and $E_{F}=\mu_{1}$.
The inset in (a) illustrates Brillouin zone with a ditrigonal pyramid
in a hexagonal setting \citep{Texture_2010_1}.
}
\end{figure}

The $C_{3v}$ crystal field reduces the $j$ multiplets according to
$j=5/2 \rightarrow 2\Gamma_{4}\oplus\Gamma_{5}\oplus\Gamma_{6}$ and
$j=3/2 \rightarrow \Gamma_{4}\oplus\Gamma_{5}\oplus\Gamma_{6}$,
where $\Gamma_{4}$ is the two-dimensional spinor irrep and $\Gamma_{5},\Gamma_{6}$ are one-dimensional complex-conjugate spinor irreps of the $C_{3v}$ double group \citep{GroupThe_1963,GroupThe_1980,GroupThe_1994}, see Appendix~\ref{Appnd3}. Even though each irrep is one dimensional in $\Gamma_{5}\oplus\Gamma_{6}$, they pair into a Kramers doublet at the $\Gamma$ point as enforced by time-reversal symmetry. These properties can be observed, for instance, for $j=3/2$ electrons, in Fig.~$\text{\ref{fig:Bulk-band-dispersion}}$(a) where band splitting happens in $\Gamma-L(H)$ directions.

To incorporate Kramers' degeneracy into our model, we include time-reversal
symmetry $T$ in the $C_{3v}$ point group. The resulting full group
is given by $\mathcal{M}_{3v}=C_{3v}+TC_{3v}$, where $\mathcal{M}_{3v}$
denotes the gray magnetic point group formed by combining $C_{3v}$
with time-reversal symmetry. In this case, the character table doubles
in terms of both the number of group elements and the number of irreps \citep{Texture_2021_4}. Each group element
$g\in C_{3v}$ acquires a time-reversed partner $Tg\in TC_{3v}$.
In the nonmagnetic $C_{3v}$ group, the irreps are given by $\Gamma \in \{A_{1},A_{2},E\}$.
We employ an orthogonal corepresentation scheme by separating time-reversal-even
and -odd operators at the group-theory level. In this case, the total
number of irreps is doubled in the magnetic group $\mathcal{M}_{3v}$:
$\Gamma \in \{\Gamma_{-},\Gamma_{+}\}$, where $\Gamma_{\pm} \in \{A_{1,\pm},A_{2,\pm},E_{\pm}\}$
and the $\pm$ sign indicates whether the irrep is even or odd under time-reversal.

We begin by obtaining
the matrix representation of the TAM  vector $\hat{\mathbf{J}}=(\hat{J}_{x},\hat{J}_{y},\hat{J}_{z})$
obeying SU(2) algebra in four- and six-dimensional representations for
$j=3/2$ and $j=5/2$ basis, respectively. The multiband normal state model Hamiltonian
takes the form $H=\sum_{\mathbf{k}}\hat{\psi}_{\mathbf{k}}^{\dagger}\hat{H}(\mathbf{k})\hat{\psi}_{\mathbf{k}}$
where $\hat{\psi}_{\mathbf{k}}$ with
\begin{align}
j &= \frac{3}{2}:\; \hat{\psi}_{\mathbf{k}}\!=\!(c_{{3}/{2}},c_{{1}/{2}},c_{-{1}/{2}},c_{-{3}/{2}})^{^{T}}, \\
j &= \frac{5}{2}:\; \hat{\psi}_{\mathbf{k}}\!=\!(c_{{5}/{2}},c_{{3}/{2}},c_{{1}/{2}},c_{-{1}/{2}},c_{-{3}/{2}},c_{-{5}/{2}})^{^{T}},
\end{align}
where $c_{m_{j},\mathbf{k}}^{\dagger}$ $(c_{m_{j},\mathbf{k}})$
denotes the fermionic creation (annihilation) operator labeled with
the 3D momentum $\mathbf{k}\in(k_{x},k_{y},k_{z})$ and magnetic quantum
number $m_{j}\in\{-j,-j+1,...,j-1,j\}$. The three-dimensional noncentrosymmetric
bulk Hamiltonian $\hat{H}(\mathbf{k})$, formulated in either the
$j=3/2$ or the $j=5/2$ representation near the $\Gamma$
point, is derived from group-theoretical analysis, see Appendix~$\ref{Appnd1}$ and Appendix~$\ref{Appnd2}$, as follows:
\begin{align}
\hat{H}(\mathbf{k})=\hat{\mathcal{H}}_{t}(\mathbf{k}) & +\hat{\mathcal{H}}_{\text{soc}}(\mathbf{k}),\label{Full_Hamiltonian_SOC}
\end{align}
with
\begin{equation}
\hat{\mathcal{H}}_{t}(\mathbf{k})=\mathcal{A}_{1}(\mathbf{k})\hat{J}_{0}+\mathcal{A}_{2}(\mathbf{k})\hat{J}_{z}^{2}\label{Full kinetic Term}
\end{equation}
where $\hat{\mathcal{H}}_{t}(-\mathbf{k})=\hat{\mathcal{H}}_{t}(\mathbf{k})$
is even-parity including momentum-dependent polynomials $\mathcal{A}_{1}(\mathbf{k})$
and $\mathcal{A}_{2}(\mathbf{k})$ representing the kinetic term and the $m_{j}$-dependent
SOC energy shift, both transforming according to the $A_{1,+}\otimes A_{1,+}$
direct product. Note that in Eq.~(\ref{Full kinetic Term}), we keep only the minimal time-reversal-even $C_{3v}$-allowed kinetic invariants, $\hat{J}_0$ and $\hat{J}_z^2$, which capture the leading band curvature and axial anisotropy near $\Gamma$. The remaining even terms, allowed by symmetry, are higher-order (in $\mathbf{k}$ and/or TAM rank). They are listed in Table~\ref{tab:BasisTables}. One needs to combine the TAM tensor matrix with the relevant basis polynomial such that the resulting term transforms according to the $A_{1,+}$ irrep. These terms can be included straightforwardly for material-specific modeling.
\begin{table}[t]
\centering{}\caption{Basis matrices and momentum dependent polynomials transforming according
to $\mathcal{M}_{3v}$ gray magnetic group symmetry. $\!\mathcal{A}_{i}(\mathbf{k})$
is defined in Eq.~$(\ref{Full kinetic Term})$ and $\mathcal{B}_{i}(\mathbf{k})=m_{1,i}k_{y}(3k_{x}^{2}\!-\!k_{y}^{2})\!+\!m_{2,i}k_{y}(3k_{x}^{2}\!-\!k_{y}^{2})k_{z}^{2}$.
Basis polynomial for doublet $E_{+}$ is $(\varrho_{i},\varpi_{i})$
where $\varrho_{i}\!=\!u_{1,i}k_{x}k_{z}\!+\!u_{2,i}(k_{x}^{2}-k_{y}^{2})\!+\!u_{3,i}(k_{x}^{2}-k_{y}^{2})k_{z}^{2}\!+\!u_{4,i}[(k_{x}^{2}-k_{y}^{2})^{2}\!-\!4k_{x}^{2}k_{y}^{2}]\!+\!u_{5,i}k_{x}k_{z}^{3}$,
and $\varpi_{i}\!=\!-u_{1,i}k_{y}k_{z}\!+\!2u_{2,i}k_{x}k_{y}\!+\!2u_{3,i}k_{x}k_{y}k_{z}^{2}\!-\!4u_{4,i}(k_{y}^{2}-k_{x}^{2})k_{y}k_{x}\!-\!u_{5,i}k_{y}k_{z}^{3}$,
where $m_{\nu,i}$ and $u_{\nu,i}$ are material dependent parameters
controlling the strength of spin-orbit energy. Leading order of TAM (momenta)
in $E_+$ and $A_{2-}$ are dipolar and quadrupolar (cubic and quadratic), respectively.}\label{tab:BasisTables}
\begin{tabular}{cccc}
\hline
\hline
 & TAM tensor & Polynomials & $j>1/2$\tabularnewline
\hline
$A_{1,+}$ & $\hat{J}_{0}$ & $\!\mathcal{A}_{i}(\mathbf{k})$ & \foreignlanguage{american}{}\tabularnewline
$A_{1,+}$ & $\hat{J}_{z}^{2},\lceil(\hat{J}_{x}^{3}-3\llbracket\hat{J}_{x}\hat{J}_{y}^{2}\rrbracket)\hat{J}_{z}\rfloor,\hat{J}_{z}^{4}$ & $\!\mathcal{A}_{i}(\mathbf{k})$ & $\checkmark$\tabularnewline
\hline
$A_{2,+}$ & $\lceil(3\llbracket\hat{J}_{y}\hat{J}_{x}^{2}\rrbracket-\hat{J}_{y}^{3})\hat{J}_{z}\rfloor$ & $k_{z}(3k_{y}k_{x}^{2}\!-\!k_{y}^{3})$ & $\checkmark$\tabularnewline
\hline
\foreignlanguage{american}{$E_{+}$} & $\hat{J}_{x}^{2}-\hat{J}_{y}^{2}$ & \foreignlanguage{american}{$\varrho_{i}(\mathbf{k})$} & $\checkmark$\tabularnewline
\foreignlanguage{american}{} & \foreignlanguage{american}{$2\lceil\hat{J}_{x}\hat{J}_{y}\rfloor$} & \foreignlanguage{american}{$\varpi_{i}(\mathbf{k})$} & $\checkmark$\tabularnewline
\hline
\foreignlanguage{american}{$E_{+}$} & \foreignlanguage{american}{$\lceil\hat{J}_{x}\hat{J}_{z}\rfloor$} & \foreignlanguage{american}{$\varrho_{i}(\mathbf{k})$} & $\checkmark$\tabularnewline
\foreignlanguage{american}{} & \foreignlanguage{american}{$\lceil\hat{J}_{y}\hat{J}_{z}\rfloor$} & \foreignlanguage{american}{$-\varpi_{i}(\mathbf{k})$} & $\checkmark$\tabularnewline
\hline
\foreignlanguage{american}{$E_{+}$} & \foreignlanguage{american}{$\lceil\hat{J}_{x}\hat{J}_{z}^{3}\rfloor$} & \foreignlanguage{american}{$\varrho_{i}(\mathbf{k})$} & $\checkmark$\tabularnewline
\foreignlanguage{american}{} & \foreignlanguage{american}{$\lceil\hat{J}_{y}\hat{J}_{z}^{3}\rfloor$} & \foreignlanguage{american}{$-\varpi_{i}(\mathbf{k})$} & $\checkmark$\tabularnewline
\hline
$A_{1,-}$ & $3\llbracket\hat{J}_{y}\hat{J}_{x}^{2}\rrbracket-\hat{J}_{y}^{3}$ & $k_{z}$ & $\checkmark$\tabularnewline
$A_{1,-}$ & $\lceil(3\llbracket\hat{J}_{y}\hat{J}_{x}^{2}\rrbracket\!-\!\hat{J}_{y}^{3})\hat{J}_{z}^{2}\rfloor$ & $k_{z}$ & $\checkmark$\tabularnewline
\hline
$A_{2,-}$ & $\hat{J}_{z}$ & $\mathcal{B}_{i}(\mathbf{k})$ & \foreignlanguage{american}{}\tabularnewline
$A_{2,-}$ & $\hat{J}_{x}^{3}\!-\!3\llbracket\hat{J}_{x}\hat{J}_{y}^{2}\rrbracket$ & $\mathcal{B}_{i}(\mathbf{k})$ & $\checkmark$\tabularnewline
$A_{2,-}$ & $\hat{J}_{z}^{3}$ & $\mathcal{B}_{i}(\mathbf{k})$ & $\checkmark$\tabularnewline
\hline
$E_{-}$ & $\hat{J}_{y}$ & $\!\varLambda_{i}(\mathbf{k})$ & \foreignlanguage{american}{}\tabularnewline
\foreignlanguage{american}{} & $\hat{J}_{x}$ & $\!\Upsilon_{i}(\mathbf{k})$ & \foreignlanguage{american}{}\tabularnewline
\hline
$E_{-}$ & $\lceil\hat{J}_{y}\hat{J}_{z}^{2}\rfloor$ & $\!\varLambda_{i}(\mathbf{k})$ & $\checkmark$\tabularnewline
\foreignlanguage{american}{} & $\lceil\hat{J}_{x}\hat{J}_{z}^{2}\rfloor$ & $\!\Upsilon_{i}(\mathbf{k})$ & $\checkmark$\tabularnewline
\hline
$E_{-}$ & $\llbracket\hat{J}_{x}\hat{J}_{y}\hat{J}_{z}\rrbracket$ & $\!\varLambda_{i}(\mathbf{k})$ & $\checkmark$\tabularnewline
\foreignlanguage{american}{} & $\frac{1}{2}\llbracket(\hat{J}_{y}^{2}-\hat{J}_{x}^{2})\hat{J}_{z}\rrbracket$ & $\!\Upsilon_{i}(\mathbf{k})$ & $\checkmark$\tabularnewline
\hline
\hline
\end{tabular}
\end{table}
In Eq.~(\ref{Full kinetic Term}), the polynomials are defined by
\begin{equation}
\mathcal{A}_{i}(\mathbf{k})=\mu_{i}+\alpha_{i,1}(k_{x}^{2}+k_{y}^{2})+\alpha_{i,2}k_{z}^{2},
\label{Ai_coef}
\end{equation}
where $\alpha_{i,1}$ and $\alpha_{i,2}$ 
specify the in-plane and out-of-plane kinetic energy, respectively, and $\mu_1$ ($\mu_2$) denotes the Fermi energy ($m_{j}$-dependent energy shift). 

In the following, we illustrate our general analysis for the non-trivial case of $j=3/2$ electrons, where $\hat{\mathbf{J}}$ are $4\times 4$ matrices. This choice allows us to still obtain analytical results. The same symmetry-based construction applies to other $j>1/2$ manifolds (including $j=5/2$); only the algebraic complexity of the explicit matrix expressions increases.


In Eq.~$(\ref{Full kinetic Term})$,
when $\mathcal{A}_{2}(\mathbf{k})\neq0$, the fourfold degeneracy
of the $j=3/2$ manifold in $\mathcal{A}_{1}(\mathbf{k})$ splits
into a pair of doubly degenerate energy bands, see Fig.~$\text{\ref{fig:Bulk-band-dispersion}}$(a),
given by
\begin{align}
E_{\mathrm{HM}}(\mathbf{k}) & =\mathcal{A}_{1}(\mathbf{k})+\frac{9}{4}\mathcal{A}_{2}(\mathbf{k}),\\
E_{\mathrm{LM}}(\mathbf{k}) & =\mathcal{A}_{1}(\mathbf{k})+\frac{1}{4}\mathcal{A}_{2}(\mathbf{k}),
\end{align}
where the labels ``heavy'' (HM) and ``light'' (LM) refer to the band
curvature (effective mass) near the $\Gamma$ point. For a dispersion
$E(\mathbf{k})$, the inverse effective-mass tensor is defined as
$(m^{-1})_{\nu\nu'}(\Gamma)=[(1/\hbar^{2})\partial^{2}E(\mathbf{k})/\partial k_{\nu}\partial k_{\nu'}]\vert_{\mathbf{k}=\Gamma}$.
Both branches are quadratic and anisotropic, and $(m^{-1})_{\nu\nu'}(\Gamma)$
is diagonal. The curvature coefficients are
\begin{align}
M_{\perp}^{\mathrm{HM}} & =\alpha_{1,1}+\frac{9}{4}\alpha_{2,1}, & M_{z}^{\mathrm{HM}} & =\alpha_{1,2}+\frac{9}{4}\alpha_{2,2},\\
M_{\perp}^{\mathrm{LM}} & =\alpha_{1,1}+\frac{1}{4}\alpha_{2,1}, & M_{z}^{\mathrm{LM}} & =\alpha_{1,2}+\frac{1}{4}\alpha_{2,2}.
\end{align}
Consequently, the effective masses are $m_{\perp}^{\lambda}=\hbar^{2}/2M_{\perp}^{\lambda}$
and $m_{z}^{\lambda}=\hbar^{2}/2M_{z}^{\lambda}$ with $\lambda\in\{\text{HM,LM}\}$. Thus, along a given direction the branch is ``heavier'' (``lighter'') if it has smaller (larger) curvature magnitude, i.e. larger (smaller) $|m_{\perp,z}^{\lambda}|$. Note that this mass-based labeling is distinct from the energetic ordering of the two branches: their energy separation is $\Delta E(\mathbf{k})=2\,\mathcal{A}_{2}(\mathbf{k})$, 
so the sign of $\mathcal{A}_{2}(\mathbf{k})$ controls which branch lies higher in energy 
at a given $\mathbf{k}$.

In Eq.~(\ref{Full_Hamiltonian_SOC}), $\hat{\mathcal{H}}_{\text{soc}}(\mathbf{k})$
denotes the odd-parity SOC term present due to broken inversion symmetry,
i.e., $\hat{\mathcal{H}}_{\text{soc}}(-\mathbf{k})=-\hat{\mathcal{H}}_{\text{soc}}(\mathbf{k})$,
transforming according to the decomposition $A_{1,+}=E_{-}\otimes E_{-}$, given explicitly by
\begin{equation}
\hat{\mathcal{H}}_{\text{soc}}(\mathbf{k})=\hat{\mathcal{H}}_{\text{MR}}(\mathbf{k})+\hat{\mathcal{H}}_{\text{HS}}(\mathbf{k}),\label{full_odd_SOC}
\end{equation}
where $\hat{\mathcal{H}}_{\text{MR}}(\mathbf{k})$ denotes the SOC
term constructed from rank-$1$ (dipolar) basis functions. We call it modified Rashba (MR). It can written as
\begin{equation}
\hat{\mathcal{H}}_{\text{MR}}(\mathbf{k})=\Upsilon_{1}(\mathbf{k})\hat{J}_{x}+\varLambda_{1}(\mathbf{k})\hat{J}_{y}.\label{MR}
\end{equation}
Additionally, $\hat{\mathcal{H}}_{\mathrm{HS}}(\mathbf{k})$ denotes the contribution
from higher-rank multipolar basis functions, given by
\begin{align}
\hat{\mathcal{H}}_{\text{HS}}(\mathbf{k}) &= \hat{\mathcal{H}}_{\text{HS}}^{(1)}(\mathbf{k})+\hat{\mathcal{H}}_{\text{HS}}^{(2)}(\mathbf{k}),\\
\hat{\mathcal{H}}_{\text{HS}}^{(1)}(\mathbf{k}) &= \Upsilon_{2}(\mathbf{k})\lceil\hat{J}_{x}\hat{J}_{z}^{2}\rfloor+\varLambda_{2}(\mathbf{k})\lceil\hat{J}_{y}\hat{J}_{z}^{2}\rfloor,\\
\hat{\mathcal{H}}_{\text{HS}}^{(2)}(\mathbf{k}) &= \frac{1}{2}\Upsilon_{3}(\mathbf{k})\llbracket(\hat{J}_{y}^{2}-\hat{J}_{x}^{2})\hat{J}_{z}\rrbracket+\varLambda_{3}(\mathbf{k})\llbracket\hat{J}_{x}\hat{J}_{y}\hat{J}_{z}\rrbracket,\label{HighSpin_SOC_odd_parity}
\end{align}
where the symmetrization operator for triple and double products is defined by $\llbracket\hat{A}\hat{B}\hat{C}\rrbracket = (1/6) (\hat{A}\hat{B}\hat{C} + \hat{C}\hat{A}\hat{B} + \hat{B}\hat{C}\hat{A} + \hat{C}\hat{B}\hat{A} + \hat{A}\hat{C}\hat{B} + \hat{B}\hat{A}\hat{C})$
and $\lceil\hat{A}\hat{B}\rfloor = (1/2) (\hat{A}\hat{B}+\hat{B}\hat{A})$, respectively.  
Both $\Upsilon_{i}(\mathbf{k})$ and $\varLambda_{i}(\mathbf{k})$, which appear in Eqs.~(\ref{MR})--(\ref{HighSpin_SOC_odd_parity}), are odd functions of momentum.  They can be written as polynomials in $\mathbf{k}$
up to the fifth order as
\begin{align}
\varLambda_{i}(\mathbf{k}) &= \gamma_{1,i}k_{x}+\gamma_{2,i}(k_{x}^{2}-k_{y}^{2})k_{z}+\gamma_{3,i}k_{x}k_{z}^{2}\nonumber \\
 &\quad{} +\gamma_{4,i}(k_{x}^{5}\!-\!10k_{x}^{3}k_{y}^{2}\!+\!5k_{x}k_{y}^{4}),\label{Lamdai}\\
\Upsilon_{i}(\mathbf{k}) &= -\gamma_{1,i}k_{y}+\gamma_{2,i}(2k_{x}\ k_{y}\ k_{z})-\gamma_{3,i}k_{y}k_{z}^{2}\nonumber \\
 &\quad{} +\gamma_{4,i}(k_{y}^{5}-\!10k_{x}^{2}k_{y}^{3}+5k_{x}^{4}k_{y}),
 \label{Upsiloni}
\end{align}
where $\gamma_{1,i}$ defines the SOC linear in momentum and $\gamma_{2(3),i}$, 
and $\gamma_{4,i}$ are cubic and quintic in momentum, respectively.
Importantly, $\hat{\mathcal{H}}_{\text{HS}}(\mathbf{k})$ is exclusive
for high TAM. This is because the squared angular momentum operators
do not result in the identity matrix, i.e., $\hat{J}_{i}^{2}\neq1$,
unlike for the $j=1/2$ electrons. Note that  $\hat{\mathcal{H}}^{(1)}_{\text{HS}}(\mathbf{k})$ and $\hat{\mathcal{H}}_{\text{MR}}(\mathbf{k})$ are identical in the $j=1/2$ basis. They reduce to the conventional Rashba SOC up to linear momenta $\hat{\mathcal{H}}_{1/2}(\mathbf{k})=k_{x}\hat{\sigma}_{y}-k_{y}\hat{\sigma}_{x}$, and $\hat{\mathcal{H}}^{(2)}_{\text{HS}}(\mathbf{k})$ vanishes.

The analogue of the spin-texture vector in the $j=1/2$
basis is the total-angular-momentum texture vector,
\begin{equation}
\mathbf{J}_{n}(\mathbf{k})=\big(\langle\hat{J}_{x}\rangle_{n\mathbf{k}},\langle\hat{J}_{y}\rangle_{n\mathbf{k}},\langle\hat{J}_{z}\rangle_{n\mathbf{k}}\big),
\end{equation}
where $\langle\hat{J}_{i}\rangle_{n\mathbf{k}}=\langle u_{n\mathbf{k}}\vert\hat{J}_{i}\vert u_{n\mathbf{k}}\rangle$
denotes the expectation value of the $i$-th component of the total
angular-momentum operator in the Bloch eigenstate $\vert u_{n\mathbf{k}}\rangle$
of band $n$ at momentum $\mathbf{k}$. If $\mathcal{A}_{2}(\mathbf{k})\neq0$,
we obtain a constant out-of-plane TAM texture, i.e., $\mathbf{J}_{\text{HM}}(\mathbf{k})=(0,0,\pm3/2)$
and $\mathbf{J}_{\text{LM}}(\mathbf{k})=(0,0,\pm1/2)$. The net TAM texture
is zero since the $A_{1+}\otimes A_{1+}$ decomposition is symmetric under
time-reversal and parity operations. 

\section{\label{sec:Symmetry-properties}Symmetry properties}

The full Hamiltonian in Eq.~$(\ref{Full_Hamiltonian_SOC})$ is invariant
under the time-reversal operation, $\hat{\mathcal{T}}\hat{H}(\mathbf{k})\hat{\mathcal{T}}^{\dagger}=\hat{H}(-\mathbf{k})$,
where $\hat{\mathcal{T}}=\exp(-i\pi\hat{J}_{y})\mathcal{K}$ is the
anti-unitary time-reversal operator with $\mathcal{K}$ denoting
complex conjugation. Additionally, $\hat{H}(\mathbf{k})$ transforms
according to the $A_{1,+}$ irrep of the $\mathcal{M}_{3v}$
magnetic point group, as implied by the covariant symmetry condition
$\hat{H}(R^{-1}\mathbf{k})=\hat{g}\hat{H}(\mathbf{k})\hat{g}^{\dagger}$,
where $\hat{g}\in\{E,2C_{3z},3\sigma_{v},2TC_{3z},3T\sigma_{v},T\}$
represents the symmetry elements, including the identity operator
$E$, the three-fold rotation around the $z$-axis $C_{3z}$, the mirror
reflections $\sigma_{v}$, and their combinations with time-reversal
symmetry $T$. Here, $R$ is a $3\times3$ orthogonal rotation matrix in
momentum space. Under threefold rotation and mirror reflection, TAM
and momentum transform as follows: 
\begin{align}
R_{3z}^{-1}\mathbf{k} & =\begin{pmatrix}\tfrac{-1}{2} & \tfrac{-\sqrt{3}}{2} & 0\\
\tfrac{\sqrt{3}}{2} & \tfrac{-1}{2} & 0\!\\
0 & 0 & 1
\end{pmatrix}\mathbf{k},\\
\sigma_{v_{1}}^{-1}\mathbf{k} & =\begin{pmatrix}+1 & 0 & 0\\
0 & -1 & 0\\
0 & 0 & +1
\end{pmatrix}\mathbf{k},\\
\hat{C}_{3z}\hat{\mathbf{J}}\hat{C}_{3z}^{-1} & =\begin{pmatrix}\tfrac{-1}{2} & \tfrac{\sqrt{3}}{2} & 0\\
\tfrac{-\sqrt{3}}{2} & \tfrac{-1}{2} & 0\\
0 & 0 & 1
\end{pmatrix}\hat{\mathbf{J}},\\
\hat{M}_{1}\hat{\mathbf{J}}\hat{M}_{1}^{-1} & =\begin{pmatrix}-1 & 0 & 0\\
0 & +1 & 0\\
0 & 0 & -1
\end{pmatrix}\hat{\mathbf{J}},
\end{align}
where $\hat{C}_{3z}=\exp(-2\pi i\hat{J}_{z}/3)$ and $\hat{M}_{1}=\exp(-i\pi\hat{J}_{y})$
denotes the matrix form in $j\in\{3/2,5/2\}$ basis for  threefold
rotation and mirror reflection operators, respectively. Under time-reversal
operation, both the TAM vector $\hat{\mathbf{J}}$ and the momentum
vector $\mathbf{k}$ reverse sign, i.e., $\hat{\mathbf{J}}\rightarrow-\hat{\mathbf{J}}$
and $\mathbf{k}\rightarrow-\mathbf{k}$. The symmetry-allowed terms
in the Hamiltonian $\hat{H}(\mathbf{k})$ are summarized in Table~\ref{tab:BasisTables},
where we list all allowed TAM and momentum bases that transform
properly under $\mathcal{M}_{3v}$. Note that we adopt the conventions of Ref.~\citep{Katzer_CT}, in which mirror reflection symmetry
implies $k_{y}\rightarrow-k_{y}$. However, for the trigonal P31m
space group, as in $\text{PtBi}_{2}$, the mirror planes are oriented
differently so that the $A_{2-}$ basis function is rotated to $k_{x}(3k_{y}^{2}\!-\!k_{x}^{2})$~\citep{SOC_2025b}. 


\section{\label{sec:Band-Basis-representation}Band Basis representation}

Without loss of generality, we represent Eq.~$\eqref{Full_Hamiltonian_SOC}$
in the eigenbasis of $\hat{\mathcal{H}}_{t}(\mathbf{k})$, where the
magnetic quantum number $m_{j}$ serves as a proper band index, i.e.,
$\hat{V}^{\dagger}\hat{\mathcal{H}}_{\text{t}}(\mathbf{k})\hat{V}=\text{diag}(E_{\text{LB}}(\mathbf{k}),E_{\text{HB}}(\mathbf{k}))$.
In this basis defined by
\begin{equation}
\hat{V}=\left(\begin{array}{cccc}
0 & 0 & 0 & 1\\
0 & 1 & 0 & 0\\
1 & 0 & 0 & 0\\
0 & 0 & 1 & 0
\end{array}\right),
\end{equation}
the full SOC Hamiltonian takes the for\foreignlanguage{american}{m
\begin{align}
\hat{\mathscr{H}}(\mathbf{k}) & =\hat{V}^{\dagger}\hat{H}(\mathbf{k})\hat{V}\nonumber\\
 & =\left(\begin{array}{cc}
\hat{H}_{\text{LM}}(\mathbf{k}) & \hat{C}(\mathbf{k})\\{}
[\hat{C}(\mathbf{k})]^{\dagger} & \hat{H}_{\text{HM}}(\mathbf{k})
\end{array}\right)\nonumber\\
 & \equiv\hat{H}_{0}(\mathbf{k})+\hat{V}_{\text{off}}(\mathbf{k}),
\end{align}
}where the first term is block-diagonal $\hat{H}_{0}(\mathbf{k})=\text{diag}(\hat{H}_{\text{LM}}(\mathbf{k}),\hat{H}_{\text{HM}}(\mathbf{k}))$
with\foreignlanguage{american}{ $\hat{H}_{\text{HM}}=E_{\text{HM}}(\mathbf{k})\hat{\sigma}_{0}$,}
and $\hat{V}_{\text{off}}$ is the off-diagonal coupling part denoting
the SOC coupling between light and heavy bands, namely interband SOC. These terms explicitly read
\begin{align}
\hat{H}_{\text{LM}}(\mathbf{k}) &= \left(\begin{array}{cc}
E_{\text{LM}}(\mathbf{k}) & g(\mathbf{k})\\{}
[g(\mathbf{k})]^{\dagger} & E_{\text{\text{LM}}}(\mathbf{k})
\end{array}\right),\label{H_LH}\\
\hat{V}_{\text{off}} (\mathbf{k}) &= \left(\begin{array}{cc}
0 & \hat{C}(\mathbf{k})\\{}
[\hat{C}(\mathbf{k})]^{\dagger} & 0
\end{array}\right),\label{Voff}
\end{align}
where the intraband SOC in the LM sector
can be written as
\begin{equation}
g(\mathbf{k}) = \mathcal{Z}_{1}^{+}(\mathbf{k})+\frac{1}{4}\mathcal{Z}_{2}^{+}(\mathbf{k})
\end{equation}
with $\mathcal{Z}_{\nu}^{\pm}(\mathbf{k})=\Upsilon_{\nu}(\mathbf{k})\pm i\varLambda_{\nu}(\mathbf{k})$.
The off-diagonal sector $\hat{C}$ in Eq.~$\text{(\ref{Voff})}$
is given by
\begin{equation}
\hat{C}(\mathbf{k})=\left(\begin{array}{cc}
\mathcal{F}(\mathbf{k}) & \mathcal{C}(\mathbf{k})\\
-[\mathcal{C}(\mathbf{k})]^{*} & [\mathcal{F}(\mathbf{k})]^{*}
\end{array}\right),
\end{equation}
where the interband SOC elements read
\begin{align}
\mathcal{F}(\mathbf{k}) &= \frac{\sqrt{3}}{2}[\mathcal{Z}_{1}^{-}(\mathbf{k})+\frac{5}{4}\mathcal{Z}_{2}^{-}(\mathbf{k})],\label{Fk}\\
\mathcal{C}(\mathbf{k}) &= -\frac{\sqrt{3}}{4}\mathcal{Z}_{3}^{-}(\mathbf{k}).\label{ck}
\end{align}
The spectrum of $\hat{\mathscr{H}}(\mathbf{k})$
is analytically solvable as
\begin{align}
\!\!\!E_{1,\pm}(\mathbf{k}) &= \pm\frac{\mathcal{D}(\mathbf{k})-|g(\mathbf{k})|}{2},\label{General_dispersion_1}\\
\!\!\!E_{2,\pm}(\mathbf{k}) &= \pm\frac{\mathcal{D}(\mathbf{k})+|g(\mathbf{k})|}{2},\label{General_dispersion_2}
\end{align}
where $\mathcal{D}=\sqrt{|g|^{2}+4A}$, $A\equiv|\mathcal{C}|^{2}+|\mathcal{F}|^{2}$
denotes the full interband SOC, and $|A|=\sqrt{AA^{*}}$. For brevity, the $\mathbf{k}$ dependence is omitted henceforth. Importantly,
the spectrum reduces to the conventional spin splitting $E_{2,\pm}(\mathbf{k})=\pm|g|$
for $j=1/2$ electrons while $E_{1,\pm}(\mathbf{k})$ vanishes. This
is because the interband SOC is absent in a two band model, i.e.,
$\mathcal{C}=\mathcal{F}=0$. 

Generally, the analysis of the spectrum
of $\hat{\mathcal{H}}_{\text{soc}}$ can be summarized as follows. A pure modified Rashba term $\hat{\mathcal{H}}_{\text{MR}}$
emerges for $\mathcal{C}=0$, $\mathcal{F}=(\sqrt{3}/2)\mathcal{Z}_{1}^{-}$,
and $g=\mathcal{Z}_{1}^{+}$ so that both intraband SOC ($g$) and interband
SOC ($\mathcal{F}$) are present, see Fig.~$\text{\ref{fig:Bulk-band-dispersion}}$(b). 
The positive branches of the spectrum are $(1/2)\sqrt{|g|^{2}+3|\mathcal{Z}_{1}^{-}|^{2}}\!\pm\!|g|$,
implying that the interband SOC results in an additional monotonic splitting.
The doubly degeneracy along the $\Gamma-A$ path is protected by the $\mathcal{M}_{3v}$
space group since $\Upsilon_{1}$ and $\varLambda_{1}$ are independent
of $k_{z}.$ However, it can be split when an additional SOC channel like $A_{1,-}\otimes A_{1,-}$ is included in the SOC Hamiltonian. Furthermore, a pure $\hat{\mathcal{H}}_{\text{HS}}^{(1)}$
emerges for $\mathcal{C}=0$ and $\mathcal{F}=(5\sqrt{3}/8)\mathcal{Z}_{2}^{-}$,
 similarly to $\hat{\mathcal{H}}_{\text{MR}}$ but with an effective
larger interband amplitude, see Fig.~$\text{\ref{fig:Bulk-band-dispersion}}$(c). Subsequently $\mathcal{D}$ increases as $\sqrt{|g|^{2}+(75/64)|\mathcal{Z}_{1}^{-}|^{2}}$
and both positive branches shift slightly upward relative to  $\hat{\mathcal{H}}_{\text{MR}}$.
A pure $\hat{\mathcal{H}}_{\text{HS}}^{(2)}$ emerges for  $g=\mathcal{F}=0$
and $\mathcal{C}=-(\sqrt{3}/4)\mathcal{Z}_{3}^{-}$, leaving only
interband SOC. Then, $\mathcal{D}=2|\mathcal{C}|$ and the spectrum
collapses into a pair of doubly degenerate levels at $E_{\pm,\pm}\in\{\pm|\mathcal{C}|,\pm|\mathcal{C}|\}$
(no interband spin splitting), as depicted in Fig.~$\text{\ref{fig:Bulk-band-dispersion}}$(d).

The spin splittings for the intraband (``intra'') and interband (``inter'') branches
are observable and given by
\begin{align}
\Delta E_{\text{intra}}^{\pm} & =\mathcal{D}\pm|g|,\\
\Delta E_{\text{inter}} & =\mathcal{D}.
\end{align}
In the weak-hybridization regime $A\ll|g|$ (compared to the intraband
SOC), $\mathcal{D}$ becomes
\begin{equation}
\mathcal{D}\approx|g|+\frac{2A}{|g|}-\frac{2A^{2}}{|g|^{3}}+O(A^{3}),
\end{equation}
giving rise to a weakly growing splitting with hybridization in the energy bands
\begin{align}
\Delta E_{\text{intra}}^{+} & \approx2|g|+\frac{2A}{|g|}-\frac{2A^{2}}{|g|^{3}}+O(A^{3}),\\
\Delta E_{\text{intra}}^{-} & \approx\frac{2A}{|g|}-\frac{2A^{2}}{|g|^{3}}+O(A^{3}).
\end{align}
However, in the strong hybridization limit $A\gg|g|$, we obtain $\mathcal{D}\approx2\sqrt{A}+|g|^{2}/4\sqrt{A}+O(|g|^{4})$.  Then,  the band splitting is dominated by interband SOC, i.e., $\Delta E_{\text{intra}}^{\pm}\approx2\sqrt{A}\pm|g|.$

\begin{figure}
\centering{}\includegraphics[scale=0.61]{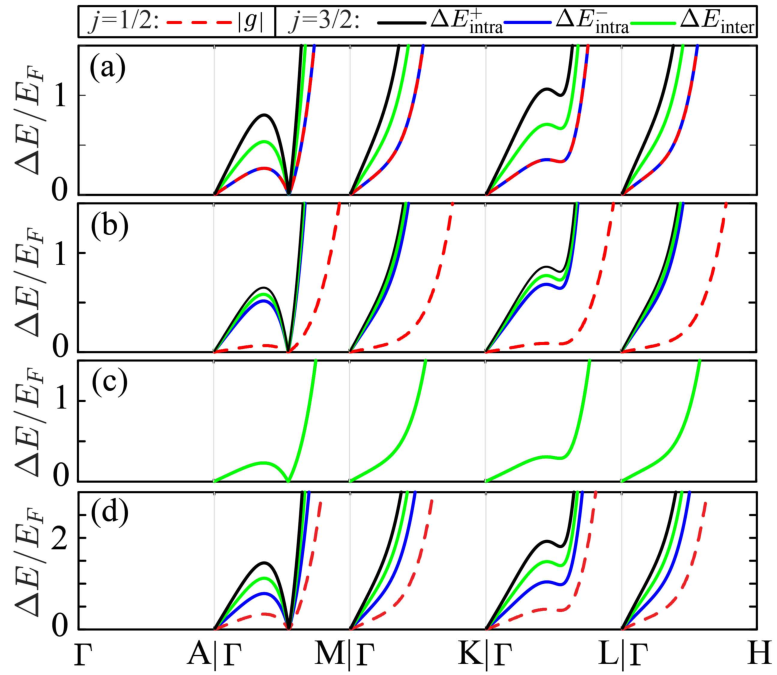}\caption{\label{SpinSplittingMagnitude}Magnitude of $E_{-}\!\otimes\!E_{-}$
intraband and interband spin splittings for $j=1/2$ and $j=3/2$
electrons near the bulk $\Gamma$ point, evaluated along a ditrigonal-pyramidal
path in the hexagonal Brillouin zone. Model parameters are fixed to $\gamma_{1,i}=\gamma_{2,i}=\gamma_{3,i}=\gamma_{4,i}=0.5\,E_{F}$.
(a) Pure modified Rashba $\hat{\mathcal{H}}_{\text{MR}}$ with $i=1$,
(b) pure $\hat{\mathcal{H}}_{\mathrm{HS}}^{(1)}$ with $i=2$, (c)
pure high-spin term $\hat{\mathcal{H}}_{\text{HS}}^{(2)}$ with $i=3$, and
(d) full SOC $\sum_{i=1}^{3}\hat{\mathcal{H}}_{i}$ with $\hat{\mathcal{H}}_{i}$
denoting pure terms in panels (a)--(c).}
\end{figure}

The magnitude of the anisotropic spin-splitting for the $j=1/2$ and $j=3/2$
states close to the $\Gamma$ point are illustrated in Fig.~$\text{\ref{SpinSplittingMagnitude}}$.
The SOC Hamiltonian includes terms up to fifth order in momentum. We plot the contributions from the modified Rashba channel $\hat{\mathcal{H}}_{\text{MR}}$ in Fig.~\ref{SpinSplittingMagnitude}(a), the high-rank SOC channels $\hat{\mathcal{H}}_{\text{HS}}^{(1,2)}$ in Figs.~\ref{SpinSplittingMagnitude}(b) and \ref{SpinSplittingMagnitude}(c),
and the full term $\hat{\mathcal{H}}_{\text{soc}}$ is illustrated in Fig.~$\text{\ref{SpinSplittingMagnitude}}$(d).
Notably, the splitting in the $j=3/2$ bands ($\Delta E_{\text{intra}}^{+}$)
exceeds that of the $j=1/2$ states; compare the black curves in Figs.~\ref{SpinSplittingMagnitude}(a) and \ref{SpinSplittingMagnitude}(b).
Although the splitting generated by $\hat{\mathcal{H}}_{\text{HS}}^{(2)}$
is purely of interband character {[}see Fig.~$\text{\ref{SpinSplittingMagnitude}}$(c){]},
its magnitude is smaller than in $\hat{\mathcal{H}}_{\text{MR}}$,
and its higher rank counterpart $\hat{\mathcal{H}}_{\text{HS}}^{(1)}$.
Importantly, along $\Gamma$-$M$, the splitting from $\hat{\mathcal{H}}_{\text{MR}}$
and $\hat{\mathcal{H}}_{\text{HS}}^{(1)}$ vanishes at a critical
momentum $\mathbf{k}_{c}$ and afterward rises sharply, due to the dominant
fifth-order terms, see Figs.~\ref{SpinSplittingMagnitude}(a) and \ref{SpinSplittingMagnitude}(b).
This can be verified analytically by setting $k_{y}=k_{x}/\sqrt{3}$ and $k_{z}=0$,
then, the component of splitting becomes $\Lambda_{1}\propto-k_{x}(4k_{x}^{2}-3)(4k_{x}^{2}+3)$
and $\Upsilon_{1}\propto-\Lambda_{1}$. Therefore, $\Lambda_{1}$
vanishes at $k_{x}=\pm\sqrt{3}/2$. Thus, the splitting 
vanishes at $\mathbf{k}_{c}=(k_{x},k_{y})=(\sqrt{3}/2,1/2)$. In addition, along $\Gamma-A$, splitting is forbidden by the $\mathcal{M}_{3v}$
point group in the $E_{-}\otimes E_{-}$ decomposition. Note that in Fig.~$\text{\ref{SpinSplittingMagnitude}}$(a),
the $j=1/2$ splitting matches that of the $j=3/2$ band $E_{1,\pm}(\mathbf{k})$
because $A=3/4|g|^{2}$ and $\mathcal{D}=2|g|$, yielding $\Delta E_{\text{intra}}^{-}=\mathcal{D}-|g|=|g|$.

Let us emphasize that interband SOC, beyond the familiar intraband Rashba splitting,
has been directly observed in Bi-based surface alloys, where hybridization
between Rashba-split bands strongly reshapes the dispersion and spin-orbital
character of surface states \citep{Interband_SOC_2012,Interband_SOC_2017}.
It leaves clear fingerprints in optical responses—for example, the
suppression of interband absorption in the persistent spin-helix regime
of Rashba--Dresselhaus systems. Therefore, it serves as a sensitive
probe of engineered SOC \citep{Interband_SOC_2013}. These observations
motivate effective models that explicitly include interband SOC terms
to faithfully capture spin-orbit-entangled bands in low-dimensional
electron systems~\citep{Interband_SOC_2022}.

\section{\label{sec:Winding-of-total}Winding of the total-angular-momentum field}

A pure modified Rashba term $\hat{\mathcal{H}}_{\mathrm{MR}}$ yields
an in-plane TAM texture whose winding number is determined by the
momentum-polynomial order. In this case, we find that the TAM texture
for a given $j$ is independent of $m_{j}$ and, in fact, identical
for all multiplets considered, $W_{j,m_{j}}=W$
where $j\in\{{1}/{2},{3}/{2},{5}/{2}\}$ and $m_{j}\in[-j,\ldots,j]$.
Therefore, the light and heavy bands share the same TAM texture. Interestingly,
once the multipolar terms are included, this universality
is lost and the heavy and light bands acquire different winding numbers.
This can be understood as a consequence of the multipolar components
associated with $\lvert m_{j}\rvert > 1/2$, which transform
under a different symmetry than the dipolar Rashba term and therefore
reshape the texture. To show this, we first discuss the TAM vorticity
phase diagram for the modified Rashba term $\hat{\mathcal{H}}_{\mathrm{MR}}$
including momentum-dependent contributions up to fifth order. This
serves as a reference against which we compare the effects of the
multipolar terms.

Importantly, in the $C_{3v}$ point group symmetry, the $z$-axis is fixed by symmetry as the unique principal $C_{3}$ rotation axis of the crystal, with all mirror planes containing this axis. Accordingly, the vorticity is defined as the winding of the in-plane (i.e., perpendicular-to-$z$) TAM texture along closed loops taken in planes normal to this symmetry axis. Any other sample orientation is handled by a coordinate rotation aligning $z$ with the crystal's $C_{3}$ axis. Moreover, $C_{3v}$ and $T$ symmetries constrain the out-of-plane TAM component to be momentum independent and forbids symmetry-allowed terms of the form $k_z \hat{J}_z$. Therefore, the reciprocal-space Skyrmion number is trivial. In this case, a 2D in-plane winding number can be nontrivial where $k_z$ is treated as a constant.

\subsection{TAM-texture for modified Rashba term $\hat{\mathcal{H}}_{\mathrm{MR}}$}

We assume $j=3/2$ and obtain the winding number of the TAM field in reciprocal space. $\hat{\mathcal{H}}_{\text{MR}}$ exhibits an in-plane TAM texture
(the out-of-plane TAM texture is vanishing) given by
\begin{equation}
\mathbf{J}_{n}(\mathbf{k})=\frac{n}{\sqrt{[\varLambda_{1}(\mathbf{k})]^{2}+[\Upsilon_{1}(\mathbf{k})]^{2}}}\left(\begin{array}{c}
\Upsilon_{1}(\mathbf{k})\\
\varLambda_{1}(\mathbf{k})\\
0
\end{array}\right),\label{TAM_texture_Pure_Rashba}
\end{equation}
where $n=m_{j}\in\{\pm{1}/{2},\pm{3}/{2}\}$ is the band
index and the helicity of each band is determined by the sign of
$m_{j}$: $\mathrm{sgn}(m_{j})=+1$ ($-1$) corresponds to clockwise
(anticlockwise) winding. $\mathbf{J}_{m_{j}}(-\mathbf{k})=-\mathbf{J}_{m_{j}}(\mathbf{k})$
is odd in momentum because both $\Upsilon_{1}(\mathbf{k})$ and $\varLambda_{1}(\mathbf{k})$
are odd and fulfills the helical property $\mathbf{J}_{-m_{j}}(\mathbf{k})=-\mathbf{J}_{m_{j}}(\mathbf{k})$.
$\hat{\mathcal{H}}_{\text{MR}}$ implies that\foreignlanguage{american}{
$(g,\mathcal{F},\mathcal{C})=(\mathcal{Z}_{1}^{+},\sqrt{3}/2\mathcal{Z}_{1}^{-},0)$.
In this case, }the dispersion in Eqs.~$\text{(\ref{General_dispersion_1})}$
and $\text{(\ref{General_dispersion_2})}$ becomes $E_{\pm1/2}=\mu_{1}\pm(1/2)|\mathcal{Z}_{1}^{+}|$
and $E_{\pm3/2}=\mu_{1}\pm(3/2)|\mathcal{Z}_{1}^{+}|$ where $|\mathcal{Z}_{1}^{+}|=\sqrt{\varLambda_{1}^{2}+\Upsilon_{1}^{2}}$\foreignlanguage{american}{.}

The winding number of the in-plane TAM texture in band $n$
around a closed contour $\mathcal{C}$ in momentum space is defined
as $W_{n}=(1/2\pi)\oint_{\mathcal{C}}d\varphi_{n,\mathbf{k}}$,
where $\varphi_{n,\mathbf{k}}$ is the local TAM angle in the $(J_{x},J_{y})$ plane. Equivalently, one may introduce the complex TAM field $\mathcal{Z}_{n}\equiv\langle\hat{J}_{x}\rangle_{n\mathbf{k}}+i\langle\hat{J}_{y}\rangle_{n\mathbf{k}}$,
whose phase satisfies $\varphi_{n,\mathbf{k}(t)}=\arg(\mathcal{Z}_{n}(t,\mathbf{q}))$
so that the winding number can be expressed in the general complex-analysis
form
\begin{align}
W_{n}(\mathbf{q}) &= \frac{1}{2\pi i}\oint_{|t|}\frac{\partial_{t}\mathcal{Z}_{n}(t,\mathbf{q})}{\mathcal{Z}_{n}(t,\mathbf{q})}dt\nonumber\\
 & =\sum_{i}N_{i}(\mathbf{q})-\sum_{i}M_{i}(\mathbf{q}),
\end{align}
where $N_{i}(\mathbf{q})$ and $M_{i}(\mathbf{q})$ denote, respectively,
the multiplicities of zeros and poles of $\mathcal{Z}_{n}(t,\mathbf{q})$
enclosed by the contour. $W_{n}(\mathbf{q})$ is proportional
to the difference of total number of zeros $N_{i}(\mathbf{q})$ and
poles $M_{i}(\mathbf{q})$ of $\mathcal{Z}_{n}(t,\mathbf{q})$ inside
the unit circle. The explicit form for the complex map can be written
as
\begin{align}
\mathcal{Z}_{n}(t,\mathbf{q}) & =\Upsilon_{1}(t,\mathbf{q})+i\Lambda_{1}(t,\mathbf{q})\nonumber\\
 & =(q_{1}+q_{2}t_{\varphi}^{3}+q_{3}t_{\varphi}^{6})/t_{\varphi},\label{complexmapFinal}
\end{align}
where $\mathbf{q}=(q_{1},q_{2},q_{3})$ is a vector of material dependent
constants defined by $q_{1}\equiv(\gamma_{1,1}+\gamma_{3,1}k_{z}^{2})|\mathbf{k}_{\perp}|$,
$q_{2}\equiv\gamma_{2,1}|\mathbf{k}_{\perp}|^{2}k_{z}$, $q_{3}\equiv\gamma_{4,1}|\mathbf{k}_{\perp}|^{5}$,
$|\mathbf{k}_{\perp}|=\sqrt{k_{x}^{2}+k_{y}^{2}}$, and $t_{\varphi}=\text{exp}(-i\varphi)$
depends on polar angle with radius $|t|=1$. Equation $\text{(\ref{complexmapFinal})}$
has a simple pole at $t_{\varphi}=0$ and six zeros in the numerators
obtained from $q_{1}+q_{2}t_{\varphi}^{3}+q_{3}t_{\varphi}^{6}=0$,
where the solutions are $t_{\varphi}\in\{t_{1},t_{2}\}$ with $t_1$ ($t_2$)
denoting fourfold (twofold) degenerate roots given by $t_{1}=\sqrt[3]{f_{\pm}(\mathbf{q})/2}$
($t_{2}=(-1)^{2/3}t_{1}$), where $f_{\pm}(\mathbf{q})=-q_{2}/q_{3}\pm[(q_{2}^{2}-4q_{1}q_{3})^{1/2}/q_{3}]$.
Since only one pole locates inside the unit circle, this results in
$\sum_{i}M_{i}(\mathbf{q})=1$. In this case, the total number of
zeros located inside the loop $|t_{\varphi}|<1$ contribute in winding
number given by
\begin{equation}
W_{n}(\mathbf{q})=\sum_{i}N_{i}(\mathbf{q})-1,
\end{equation}
where the solutions of inequalities $t_{1}t_{1}^{*}<1$ and $t_{2}t_{2}^{*}<1$
defines unique regions of $W_{n}(\mathbf{q})$.
\begin{figure}
\begin{centering}
\includegraphics[scale=0.54]{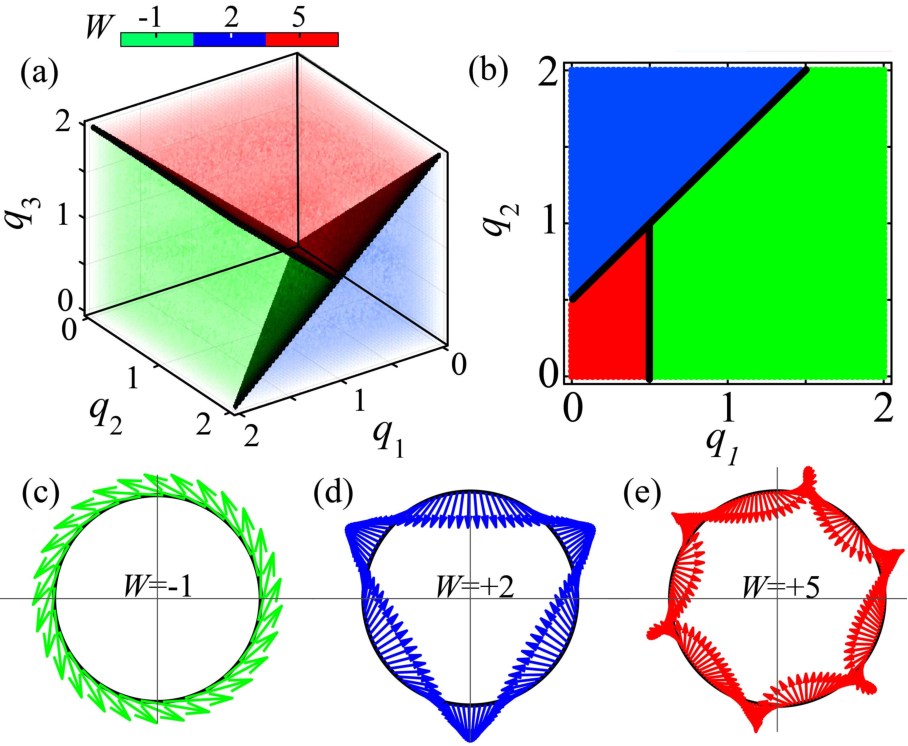}
\par\end{centering}
\caption{\label{fig:Phase Diagram-1}Phase diagrams of total-angular-momentum
vorticity as function of (a) $\mathbf{q}=(q_{1},q_{2},q_{3})$ and
(b) $\mathbf{q}=(q_{1},q_{2},0.5)$. Polar plots of the $j$ texture
for (c) single winding, $W=-1$, at $\mathbf{q}=(1.5,0,0.5)$; (d)
twofold winding, $W=2$, at $\mathbf{q}=(0.5,1.5,0.5)$; and (e) fivefold
winding, $W=5$, at $\mathbf{q}=(0.2,0.2,0.5)$. The color bar indicates
the winding number. Black planes and lines in panels (a) and (b) denote boundaries
of regions with different winding number.}
\end{figure}

The outcome is presented in the generic 3D phase diagram in Fig.~\ref{fig:Phase Diagram-1}(a). A two-dimensional (2D) slice of the phase diagram at $q_3=0.5$ is shown in Fig.~\ref{fig:Phase Diagram-1}(b). There are three distinct phases, where the TAM texture exhibits the conventional Rashba vortex (green). In this case, the TAM texture winds once around the origin, as illustrated in its polar form in
Fig.~\ref{fig:Phase Diagram-1}(c). The other regions correspond
to unconventional vortices, where the TAM texture exhibits twofold
and fivefold windings, marked in blue and red in Figs.~\ref{fig:Phase Diagram-1}(d)
and \ref{fig:Phase Diagram-1}(e), respectively. The black surfaces
indicate the phase boundaries, obtained from the conditions $t_{1}t_{1}^{*}=1$
and $t_{2}t_{2}^{*}=1$, where the zeros lie on the circumference
of the unit circle.

Importantly, in photoemission spectroscopy,
the signature of the fivefold winding can be detected as (i) the warping of
constant-energy contours near the Fermi energy and (ii) nontrivial
TAM winding around the $\Gamma$ point, either on a two-dimensional
surface or in the three-dimensional bulk \citep{Hexagonal_2010,Hexagonal_2011}.
In the latter case, larger constant-energy contours become accessible when higher-energy photons are incident on the sample or the chemical potential is changed.

\begin{figure}[t]
\begin{centering}
\includegraphics[width=0.9\linewidth]{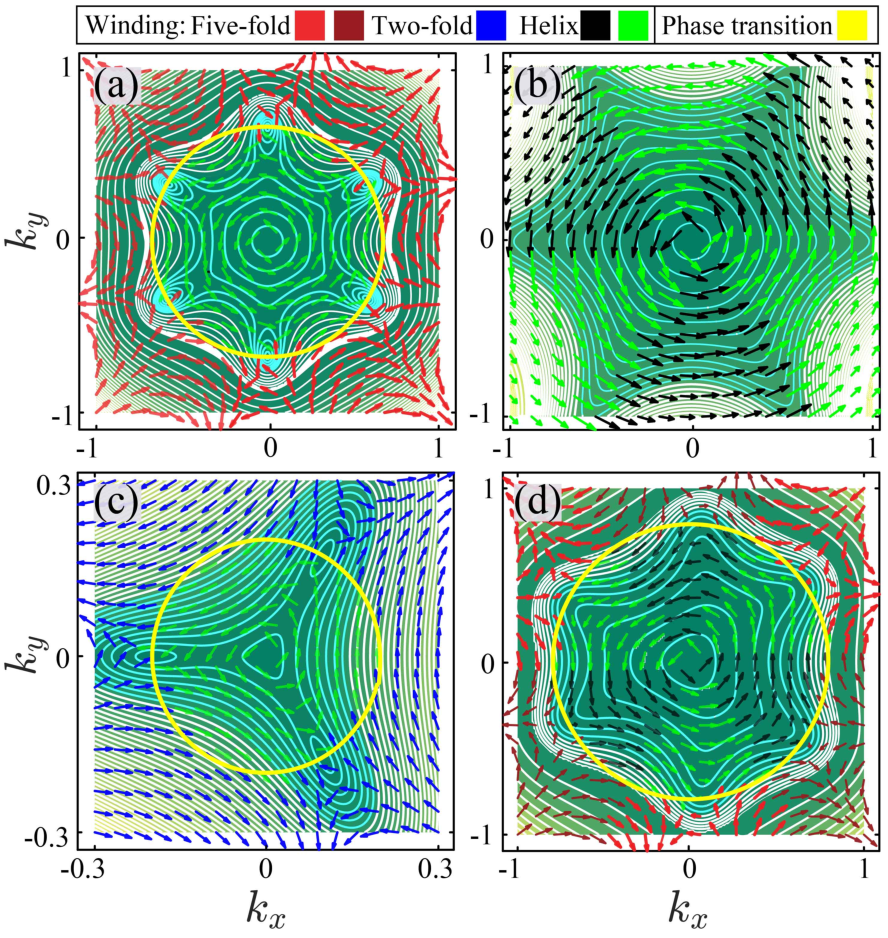}
\par\end{centering}
\caption{\label{fig:2D-constant-energy-contours-1}Two-dimensional constant-energy contours with TAM texture in $j=3/2$ basis for light-mass band, i.e., $|3/2,-1/2\rangle$. (a) Linear-quintic
mixture $E_{-}\!\otimes\!E_{-}$ with $\gamma_{1,1}/E_{F}=0.3$ and
$\gamma_{4,1}/E_{F}=1.5$, (b) linear-cubic mixture $A_{2-}\!\otimes\!A_{2-}+E_{-}\!\otimes\!E_{-}$
for slice $k_{z}=0.5$ with $\gamma_{1,1}/E_{F}=1$, $m_{1,1}/E_{F}=1.5$,
and $m_{2,1}/E_{F}=1.2$, (c) linear-cubic mixture $E_{-}\!\otimes\!E_{-}$
with $\gamma_{1,1}/E_{F}=0.3$ and $\gamma_{2,1}/E_{F}=1.5$, and (d)
full term with $\gamma_{\nu,1}/E_{F}=m_{2,1}/E_{F}=0.5$ for $\nu\in\{1,2,3\}$,
$m_{1,1}/E_{F}=1$ and $k_{z}=1$. The yellow solid circles denote the phase transition boundaries where the winding changes. The out-of-plane component of TAM texture is marked in dark and light colors. The heavy-mass
bands carry similar information. 
for other helical energy branch $W\rightarrow-W$. Other parameters are set to zero.}
\end{figure}

In Fig.~\ref{fig:2D-constant-energy-contours-1}, we show constant-energy
contours of the solutions $E_{-1/2}(\mathbf{k}_{\perp})=E_{\text{c}}/\mu_{1}$
for $E_{\text{c}}/\mu_{1}\in[-1,1]$, for a mixture of SOC terms
together with the corresponding TAM texture in the $\Gamma$-$M$-$K$ plane
denoted by momentum $\mathbf{k}_{\perp}\equiv(k_{x},k_{y},0)$. When
a linear-quintic mixture of SOC is dominant, the dispersion takes
the form
\begin{equation}
E_{n,\pm} = \mu_{1} \pm n |\mathbf{k}_{\perp}| \mathcal{G}(\varphi),
\label{AngleDispersion}
\end{equation}
where $\mathcal{G}(\varphi)$ denotes the angle dependent part of spin split dispersion given by
\begin{equation}
\mathcal{G}(\varphi) = \sqrt{\gamma_{1,1}^{2} + \gamma_{4,1}^{2}|\mathbf{k}_{\perp}|^{8} + 2\gamma_{1,1}\gamma_{4,1}|\mathbf{k}_{\perp}|^{4} \cos(6\varphi)}. \label{Dispersion_polar_dep1}
\end{equation}
It develops hexafoil energy contours for $E_{\text{c}}\geqslant\pm\mu_{1}$
due to the $\cos(6\varphi)$, shown in Fig\@.~$\text{\ref{fig:2D-constant-energy-contours-1}}$(a).
Energy contours are distributed as a central circular pocket surrounded
by six lobes. These contours evolve from an almost circular contour
at $E\approx E_{F}$ into a concave hexagon. This can be understood
that in the vicinity of the $\Gamma$ point (i.e., for $|\mathbf{k}_{\perp}|\!\ll\!1$),
the linear SOC term $\propto\mathbf{k}_{\perp}$ dominates over the
quintic term $\propto|\mathbf{k}_{\perp}|^{5}$, yielding nearly circular
constant-energy contours. In this case, the TAM texture exhibits a
conventional helix, see green arrows in Fig\@.~$\text{\ref{fig:2D-constant-energy-contours-1}}$(a).
Note that the TAM texture of a pure linear $\gamma_{1,1}$ (cubic
$\gamma_{3,1}$) SOC has a polar dependence while cubic $\gamma_{2,1}$
and quintic $\gamma_{4,1}$ has an additional radial part, as listed
in Table~\ref{tab:pure Winding-of-spin}. 
\begin{table}
\centering{}\caption{\label{tab:pure Winding-of-spin}Winding of total-angular-momentum
texture for pure odd-parity spin-orbit coupling terms $E_{-}\otimes E_{-}$
up to fifth order of momentum preserving time-reversal and $C_{3v}$
point group symmetries. Coefficients are $v_{1,3,4}=\text{sgn}(\gamma_{1(3)[4],1})$
and $v_{2}=\text{sgn}(k_{z}\gamma_{2,1})$.}
\begin{tabular}{cccc}
\hline 
\hline 
$\neq0$ & $\mathbf{k}$ & $\mathbf{J}_{n}(\varphi)$ & $|W_n|$\tabularnewline
\hline 
$\gamma_{1,1}$ & Linear & $v_{1}\mathbf{e}_{\varphi}$ & $1$\tabularnewline
\hline 
$\gamma_{2,1}$ & Cubic & $v_{2}\text{sin}(3\varphi)\mathbf{e}_{r}\!+\!v_{2}\text{cos}(3\varphi)\mathbf{e}_{\varphi}$ & $2$\tabularnewline
\hline 
$\gamma_{3,1}$ & Cubic & $v_{3}\mathbf{e}_{\varphi}$ & $1$\tabularnewline
\hline 
$\gamma_{4,1}$ & Quintic & $v_{4}\text{sin}(6\varphi)\mathbf{e}_{r}\!+\!v_{4}\text{cos}(6\varphi)\mathbf{e}_{\varphi}$ & $5$\tabularnewline
\hline 
\hline 
\end{tabular}
\end{table}

At larger momenta ($|\mathbf{k}_{\perp}|\!\gtrsim\!1$), the fifth-order
SOC becomes dominant and the contours evolve into a concave hexagon.
For $E>E_{c}$, the central pocket disappears and the six lobes merge
into a simple hexagon signaling a highly suppressed Rashba contribution.
Interestingly, the TAM texture undergoes a nontrivial transition away
from the $\Gamma$ point exhibiting quintic vortex, as illustrated
by red arrows in Fig\@.~$\text{\ref{fig:2D-constant-energy-contours-1}}$(a).
The region with a conventional TAM vortex, characterized by $W_{n}(|\mathbf{k}_{\perp}|)=-1$,
is separated from the quintic vorticity phase, $W_{n}(|\mathbf{k}_{\perp}|)=5$,
by a phase transition boundary, shown as a solid yellow line in Fig.~\ref{fig:2D-constant-energy-contours-1}(a),
across which the winding number undergoes a drastic jump. 
 
The dependence of the winding number on the in-plane momentum can
be understood as follows. When a linear-quintic SOC is favored, the zeros
of the complex map $\mathcal{Z}_{n}(t)=0$ are determined by $\mathcal{Z}_{n}(t)=\gamma_{1,1}+\gamma_{4,1}|\mathbf{k}_{\perp}|^{4}t^{6}$.
Thus, for given parameter values $\gamma_{1,1}$ and $\gamma_{4,1}$,
the TAM vortex depends on the magnitude of $|\mathbf{k}_{\perp}|$.
Even if the fifth-order coefficient is weak, its effect can be enhanced
at $|\mathbf{k}_{\perp}|>1$, resulting in a hexafoil SOC with a quintic
TAM vortex. We note that, for pure modified Rashba SOC, the winding number does
not depend on the in-plane momentum.

Up to the rank-1 (dipolar) correction, the SOC energy takes the form
\begin{equation}
\mathcal{H}_{\text{SOC}}(\mathbf{k})=\hat{\mathcal{H}}_{\mathrm{MR}}+\mathcal{B}_{1}(\mathbf{k})\hat{J}_z,\label{HelixwarpingHamil}
\end{equation}
where the second term arises from the $A_{2,-}\otimes A_{2,-}$ decomposition, with $\mathcal{B}_{1}(\mathbf{k})$ and $\hat{J}_{z}$ providing the momentum and TAM basis functions, respectively, each transforming according to the $A_{2,-}$ irrep and the momentum basis takes the form
\begin{equation}
\mathcal{B}_{1}(\mathbf{k})=m_{1,1}k_{y}(3k_{x}^{2}\!-\!k_{y}^{2})\!+\!m_{2,1}k_{y}(3k_{x}^{2}\!-\!k_{y}^{2})k_{z}^{2},
\end{equation}
where $m_{1(2),1}$ is a material-dependent coefficient that sets the strength of the axial Zeeman-type SOC term. 
Note that, after projecting Eq.~\eqref{HelixwarpingHamil} onto the effective spin-$1/2$ sector, it reduces to the familiar hexagonal-warping SOC of the surface states in the topological insulator Bi$_2$Te$_3$~\citep{SOC_2009b,Hexagonal_2010}.

The leading order in $A_{2,-}\otimes A_{2,-}$ decomposition gives rise to an out-of-plane spin texture. Either with $\gamma_{1,1},m_{1}\neq0$
or $\gamma_{1,1},\gamma_{2,1}\neq0$ hexagonal and trigonal warping
emerge \citep{SOC_2009a,SOC_2009b}, as shown in Fig.~\ref{fig:2D-constant-energy-contours-1}(b)
and Fig.~\ref{fig:2D-constant-energy-contours-1}(c), respectively.
The former (latter) SOC yields both in-plane and out-of-plane (a purely
in-plane) TAM texture. Importantly, the hexagonal warping exhibits
conventional TAM vortex $W_{n}(k_{z})=-1$ in the entire Brillouin
zone (2D surface or 3D bulk). However, in the limit where SOC exhibits
trigonal warping effect, we observe that the circular Rashba contours
in the vicinity of $\Gamma$ point evolves into a rotated boomerang. This happens since the angle-dependent spin splitting dispersion in Eq.~\eqref{AngleDispersion}, takes a $\cos(3\theta)$ form given by
\begin{equation}
\mathcal{G}(\varphi) = \sqrt{\gamma_{1,1}^{2} + |\mathbf{k}_{\perp}|k_{z} [\gamma_{2,1}^{2}|\mathbf{k}_{\perp}|k_{z}+2\gamma_{1,1}\gamma_{2,1}\cos(3\varphi)]}. \label{Dispersion_polar_dep2}
\end{equation}
The TAM texture winds twice ($W_{n}=2$) around the origin when $E>E_{c}$
at $|\mathbf{k}|\!\gtrsim\!k_{c}$, see Fig.~$\text{\ref{fig:2D-constant-energy-contours-1}}$(c).
In general, when all SOC terms contribute comparably, the constant-energy
contours evolve into a mixed trigonal-hexagonal form, as shown in Fig.~\ref{fig:2D-constant-energy-contours-1}(d).

Note that for pure modified Rashba SOC contributions, the
constant-energy contours are circular because the dispersion depends
only on the radial magnitude of the TAM texture. Any dependence on
the polar angle $\varphi$ appears only in the TAM-texture field,
not in the energy. However, the polar angle $\varphi$ becomes important
in the spectrum when SOC is mixed.
\begin{figure}
\begin{centering}
\includegraphics[scale=0.8]{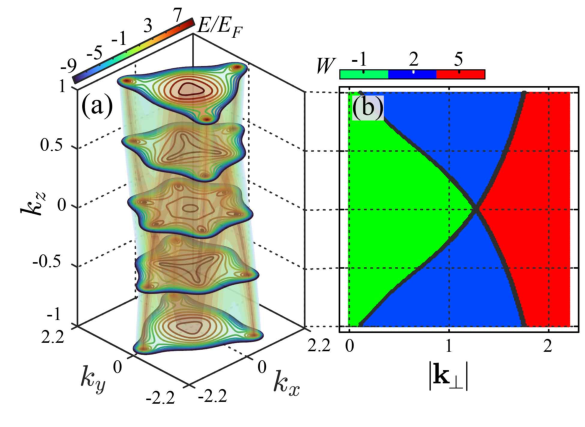}
\par\end{centering}
\caption{\label{3D bulk energy spectra-1}(a) Three-dimensional bulk spectrum
with constant-energy contours and
(b) corresponding vorticity phase diagram. Both panels are evaluated
for the light mass band with positive helicity, i.e., $|3/2,-1/2\rangle$.
Model parameters: $(\gamma_{1,1},\gamma_{2,1},\gamma_{3,1},\gamma_{4,1})/E_{F}=(-0.5,-1,0.4,0.2)$
and $|\mathbf{k}_{\perp}|=(k_{x}^{2}+k_{y}^{2})^{1/2}$. In panel
(a), the constant-energy contours are taken in the range $E/E_{F}\in[-1,1]$.
Heavy-mass bands $\lvert3/2,\pm3/2\rangle$ carry similar information
to the light-mass bands $\lvert3/2,\pm1/2\rangle$. Moreover, $W\rightarrow-W$
for opposite helicity.}
\end{figure}

Importantly, the vorticity phase diagram depends on the out-of-plane
momentum $k_{z}$. Consequently, the TAM-texture vorticity can change
dramatically as $k_{z}$ increases and states probe deeper into the
three-dimensional bulk. As an illustration, Fig.~\ref{3D bulk energy spectra-1}
shows the bulk spectrum (constant-energy contours at $k_{z}\in\{0,\pm0.5,\pm1\}$)
for a given set of model parameters together with the corresponding
vorticity phase diagram. Close to the $\Gamma$ point at $k_{z}=0$,
the constant-energy contours are hexagonal, see Fig.~\ref{3D bulk energy spectra-1}(a).
In this case, the vorticity phase diagram exhibits a conventional
Rashba vortex with $W_{n}=-1$ at small in-plane momentum $|\mathbf{k}_{\perp}|$
(marked in green) and an unconventional quintic vortex with $W_{n}=5$
at larger $|\mathbf{k}_{\perp}|$, marked in red. For larger $|k_{z}|$,
the system undergoes a vorticity phase transition where the TAM field
acquires a double winding $(\ensuremath{W_{n}=2})$, highlighted in
blue in Fig.~\ref{3D bulk energy spectra-1}(b). As $|k_{z}|$
increases further, this $W_{n}=2$ region expands while the conventional
$(\ensuremath{W_{n}=-1})$ regions shrink. Concurrently, the constant-energy
contours evolve from hexagonal—with a central circular pocket—to a
trigonal shape. By $k_{z}=\pm1$, the conventional Rashba-winding
region has almost disappeared.

\subsection{TAM-texture for multipolar SOC}

When multipolar SOC terms are included in the full Hamiltonian, they
induce anisotropy in the heavy-mass band. Consequently, the energy
bands can acquire different TAM vorticities. To show this, we assume
that the SOC includes the terms $A_{1,-} \otimes A_{1,-} + E_{-} \otimes E_{-}$.
In this case, the Hamiltonian includes a higher-rank multipolar contribution
given by
\begin{align}
\hat{\mathcal{H}}_{\text{soc}}(\mathbf{k}) &= \Upsilon_{1}(\mathbf{k})\hat{J}_{x}+\varLambda_{1}(\mathbf{k})\hat{J}_{y}\nonumber\\
 &\quad{} +\Upsilon_{2}(\mathbf{k})\lceil\hat{J}_{x}\hat{J}_{z}^{2}\rfloor+\varLambda_{2}(\mathbf{k})\lceil\hat{J}_{y}\hat{J}_{z}^{2}\rfloor\nonumber \\
 &\quad{} +b_{1}k_{z}(3\llbracket\hat{J}_{y}\hat{J}_{x}^{2}\rrbracket\!-\!\hat{J}_{y}^{3}), \label{eq:HighWingHam}
\end{align}
where the first and second lines are the SOC contributions from $\hat{\mathcal{H}}_{\text{MR}}(\mathbf{k})$
and $\hat{\mathcal{H}}_{\text{HS}}^{(1)}(\mathbf{k})$, respectively,
and the third line gives the $A_{1,-}\!\otimes\!A_{1,-}$ term with
strength $b_{1}$. 

\begin{figure}
\begin{centering}
\includegraphics[scale=0.625]{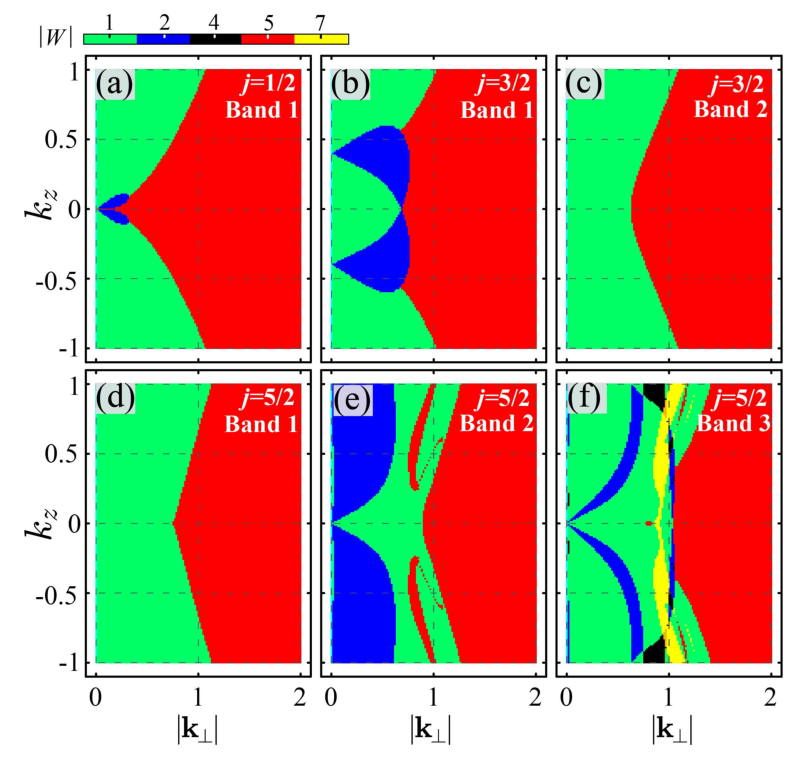}
\par\end{centering}
\caption{\label{fig:FullMultipolarPhase}Total-angular-momentum-vorticity phase
diagram for multipolar spin-orbit coupling $A_{1,-}\!\otimes\!A_{1,-}+E_{-}\!\otimes\!E_{-}$
in (a) $j=1/2$, (b), (c) $j=3/2$ and (d), (e), (f) $j=5/2$ basis, shown
for one (two) {[}three{]} lowest-energy bands, respectively. Remaining bands exhibit identical texture with opposite winding 
sign. Color bar indicates winding number of $\mathbf{J}_{n}(\mathbf{k})=\big(\langle\hat{J}_{x}\rangle_{n\mathbf{k}},\langle\hat{J}_{y}\rangle_{n\mathbf{k}}\big)$
where $n\in\{1,2,3\}$ is band index. Model parameters for multipolar
terms are $(\gamma_{1,2},\gamma_{2,2},\gamma_{3,2},\gamma_{4,2})/E_{F} = ({1}/{2},{1}/{2},{1}/{2},{1}/{2})$
and $b_{1}=0.2\,E_{F}$. Model parameters for modified Rashba term is
the same as those in Fig.~$\text{\ref{3D bulk energy spectra-1}}$.}
\end{figure}

In this case, the TAM texture no longer obeys the form given in Eq.~(\ref{TAM_texture_Pure_Rashba})
and becomes anisotropic. Consequently, each energy band exhibits a
different TAM texture. In Fig.~\ref{fig:FullMultipolarPhase},
we present the vorticity phase diagram for $j \in \{1/2,\,3/2,\,5/2\}$
electrons, obtained from the winding of the TAM texture vector $\mathbf{J}_{n}(\mathbf{k})=\big(\langle\hat{J}_{x}\rangle_{n\mathbf{k}},\langle\hat{J}_{y}\rangle_{n\mathbf{k}}\big)$
where the averages are taken over the Bloch states of Eq.~(\ref{eq:HighWingHam}).

For $j=1/2$ electrons, we consider the lowest band, while for $j=3/2$
and $j=5/2$ electrons we consider the two and three lowest bands,
respectively. The other helical branches have the opposite sign of
winding and are not shown. When SOC is purely modified-Rashba form [ignoring
the second and third lines in Eq.~(\ref{eq:HighWingHam})], the
TAM phase diagram is identical for all $j$ degrees of freedom, see
Fig.~\ref{3D bulk energy spectra-1}(b). Therefore, light-mass
electrons with $|m_{j}|=1/2$ and heavy-mass electrons with $|m_{j}|=3/2,\,5/2$
are indistinguishable.

However, the multipolar energy corrections given by the second and
third lines in Eq.~(\ref{eq:HighWingHam}) induce anisotropic TAM
textures in both the light- and heavy-mass bands. This is evident
by comparing the vorticity phase diagrams for the lowest bands of
$j=1/2$, $j=3/2$, and $j=5/2$ electrons, see Fig.~\ref{fig:FullMultipolarPhase}.
For $j=1/2$ electrons, the quintic vorticity region, marked in red, occupies a large portion of the phase diagram, whereas the twofold-winding region is confined to a small area in the vicinity of the bulk $\Gamma$ point, see Fig.~\ref{fig:FullMultipolarPhase}(a).
In addition, the linear Rashba regime (marked in green) emerges with
increasing $k_{z}$, where the TAM texture exhibits a helical form.

Interestingly, in the $j=3/2$ counterpart shown in Fig.~\ref{fig:FullMultipolarPhase}(b),
the regions of twofold-winding are enhanced and become larger (marked
in blue) in band $1$. These regions disappear for band 2, shown in Fig.~\ref{fig:FullMultipolarPhase}(c).

Band $1$ for $j=5/2$ electrons exhibits linear-quintic SOC, see Fig.~\ref{fig:FullMultipolarPhase}(d). The corresponding phase diagram is almost similar to band $2$ of $j=3/2$ electrons. Interestingly,  the anisotropy
becomes more pronounced in the second and
third lowest bands of $j=5/2$ electrons. Band $2$ evolves into a linear-cubic-quintic admixture. It has the largest area for twofold-winding in the phase diagram, see Fig.~\ref{fig:FullMultipolarPhase}(e). Importantly, the third lowest energy band, depicted in Fig.~\ref{fig:FullMultipolarPhase}(f),
exhibits exotic phases where the winding number becomes $|W_{3}^{(j=5/2)}|\in\{4,7\}$. In this case, the effective SOC energy depends on momenta up to the seventh order.

\section{\label{sec:Current-induced-spin-polarizatio}Current-induced spin
polarization}

A key manifestation of inversion-symmetry breaking is the Edelstein
effect \citep{Edel_1990}, i.e., the generation of a nonequilibrium spin (or more generally
TAM) polarization by a dc charge current and, conversely, of a
charge current by a spin imbalance \citep{Edel_2014,Edel_2020_2}.
The Edelstein effect is sensitive to both the underlying
texture and the detailed Fermi-surface geometry. It has been explored
in Weyl semimetals \citep{Edel_2018}, noncentrosymmetric antiferromagnets
\citep{Edel_2019}, superconducting interfaces \citep{Edel_2020_1},
and two-dimensional materials and Dirac systems \citep{Edel_2020_3,Edel_2021,Edel_2023}.
In systems where the low-energy states belong to high-$j$ multiplets,
it is thus essential to describe Edelstein physics in terms of $\langle\hat{\mathbf{J}}\rangle$
rather than $\langle\hat{\boldsymbol{\sigma}}\rangle$ and to relate
the magnitude and anisotropy of the response directly to the structure
and winding of the TAM texture on the Fermi surface. In this case,
we obtain in the following a counterpart of the Edelstein formalism in TAM
representation akin to the spin-$1/2$ basis \citep{Edel_1990,Edel_2011,Edel_2025_1}. 

Considering an electric field along the $x$-axis, i.e., $\mathbf{E}=(\mathrm{E}_{x},0)$,
it induces a positive spin accumulation perpendicular to the field
direction, specifically along the $y$-axis. A linear response to
the field, under the Boltzmann approximation, leads to a net finite
TAM average $\langle J_{\alpha}^{(j)}\rangle$ with $j\in\{1/2,3/2,5/2\}$,
given by $\langle J_{\alpha}^{(j)}\rangle=\sum_{\beta}\chi_{\alpha\beta}^{(j)}\mathrm{E}_{\beta}$
where $\chi_{\alpha\beta}^{(j)}$ denotes the $3\times2$ Edelstein
susceptibility tensor for angular momentum $j$, $\alpha\in\{x,y,z\}$,
and $\beta\in\{x,y\}$. Here, the components of the tensor are given
by \citep{Edel_2011,Edel_2018,Roy2022,Edel_2025_1}
\begin{equation}
\chi_{\alpha\beta}^{(j)} = e\tau\!\sum_{n=1}^{2j+1}\!\int\!\frac{d^{2}\mathbf{k}_{\perp}}{(2\pi)^{2}}\:
  J_{n}^{(\alpha)}(\mathbf{k}_{\perp})\, v_{n}^{(\beta)}(\mathbf{k}_{\perp})\,
  \mathscr{F}_{n,\mathbf{k}_{\perp}}, \label{Edelstein Tensor-1}
\end{equation}
where $\mathbf{k}_{\perp}=(k_{x},k_{y},0)$ is the in-plane momentum,
and the function $\mathscr{F}_{n,\mathbf{k}_{\perp}}=-\partial f_{0}/\partial E$
denotes the thermal broadening kernel of the linear response, $k_{B}$ is the Boltzmann constant, $T$ is temperature, $e$ and $\tau$ denote
carrier charge and momentum relaxation time, respectively. The TAM texture component
is $\mathbf{J}_{n}^{(\alpha)}(\mathbf{k}_{\perp})=\langle n,\mathbf{k}_{\perp}|\hat{J}_{\alpha}|n,\mathbf{k}_{\perp}\rangle$,
and $v_{n}^{(\beta)}(\mathbf{k}_{\perp}) = \langle n,\mathbf{k}_{\perp}| {\partial\hat{H}}/{\partial k_{\beta}} |n,\mathbf{k}_{\perp}\rangle$
denotes the group velocity for the $n$-th energy band with the eigenbasis
$|n,\mathbf{k}_{\perp}\rangle$. 

Up to linear order in the momentum-dependent
SOC and for an isotropic Fermi surface, the components of the Edelstein
tensor satisfy $\chi_{xy}=-\chi_{yx}$ and $\chi_{xx}=\chi_{yy}=0$.
The vanishing of the diagonal components follows from the angular
integration in polar coordinates, $\chi_{xx}\propto\int_{0}^{2\pi}\sin\theta\cos\theta\,d\theta=0$.
At zero temperature, the Edelstein susceptibility $\chi_{\alpha\beta}^{(j)}/e\tau$
is determined solely by the Fermi contours (FCs), since $\mathscr{F}=\delta\bigl(E_{\nu}-\mu_{c}\bigr)$
where $\mu_{c}=E_{F}$. Consequently, one can write~\citep{Edel_2011}
\begin{equation}
\chi_{xy}^{(j)}/\mathrm{c}_{0}=\sum_{n}\oint_{\mathrm{FC}_{n}}ds_{\mathrm{FC}_{n}}\;J_{n}^{(x)}(\mathbf{k}_{\perp})\,\hat{v}_{n}^{(y)}(\mathbf{k}_{\perp}),
\end{equation}
where $\mathrm{c}_{0}\equiv e\tau/(2\pi)^{2}$, $\hat{v}_{n}^{(y)}=v_{n}^{(y)}/|\mathbf{v}_{n}|$
is the $y$-component of the unit velocity vector, taking values $\hat{v}_{n}^{(y)}\in[-1,1]$,
$\mathbf{v}_{n}=(v_{n}^{(x)},v_{n}^{(y)})$ is the 2D group-velocity
vector, and $ds_{\mathrm{FC}_{n}}$ is the arc length element along
the $n$-th Fermi contour. For a circular Fermi surface, we have $ds_{\mathrm{FC}_{n}}=|\mathbf{k}_{F}^{n}|\,d\theta$.
Thus, the Edelstein susceptibility measures the net TAM polarization
along the $y$-direction carried by states at the Fermi level. The
corresponding imbalance of TAM-polarized carriers can couple to the
electric field and is transported along the group-velocity direction
\citep{Edel_2018,Edel_2020_1}. 

\begin{figure}[t]
\centering{}\includegraphics[scale=0.625]{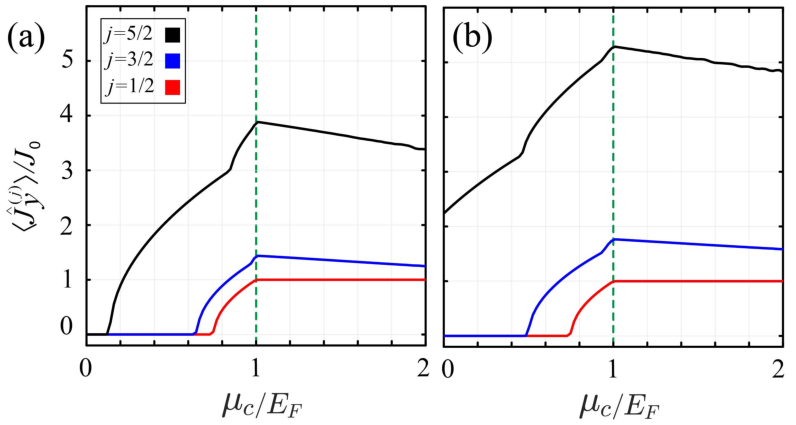}\caption{\label{fig:Edelstein_effect}Edelstein effect for total angular momenta
$j\in\{{1}/{2},{3}/{2},{5}/{2}\}$ in a non-centrosymmetric
spin-orbit coupled model Hamiltonian in Eq.~$\text{(\ref{Full_Hamiltonian_SOC})}$
that preserves time-reversal and $C_{3v}$ point group. Perpendicular
total-angular-momentum expectation value $\langle \hat{J}_{y}^{(j)}\rangle$ induced
by an external electric field $\mathbf{E}=(1,0)$ for (a) a pure modified
Rashba SOC $(\tilde{\gamma}_{1},\tilde{\gamma}_{2})=(-1,0)$ and (b)
an admixture of both SOC terms with $(\tilde{\gamma}_{1},\tilde{\gamma}_{2})=(-1,-0.3)$,
where $\tilde{\gamma}_{i}\equiv\gamma_{1,i}/E_{F}$. Model parameters for even-parity sector $\hat{\mathcal{H}}_{t}$ are
$(\tilde{\alpha}_{1},\tilde{\alpha}_{2},\tilde{\mu}_{2},\tilde{\alpha}_{1}^{\prime},\tilde{\alpha}_{2}^{\prime})=(1,0,0,1,0)$
where $\tilde{\alpha}_{i}\equiv\alpha_{1,i}/E_{F}$ and $\tilde{\alpha}_{i}^{\prime}\equiv\alpha_{2,i}/E_{F}$.
Other parameters are $J_{0}=N_{0}\text{max}(|\langle \hat{J}_{y}^{(1/2}\rangle|)$,
$\chi_{0}=N_{0}\, \text{max}(|\langle\chi_{xy}^{(1/2}\rangle|)$ where
$N_{0}=e\tau/(2\pi)^{2}$ and $E_{F}=\mu_{1}$. In the numerics, we replace
the Dirac delta function with a Gaussian function $\delta(E_{n,\mathbf{k}_{\perp}}-\mu_{c}) \equiv ({1}/{\Omega\sqrt{2\pi}})\, \exp[-(E_{n,\mathbf{k}_{\perp}}-\mu_{c})^{2}/\Omega^2]$
with broadening factor $\Omega\approx7.4\times10^{-3}$.}
\end{figure}

In Fig.~\ref{fig:Edelstein_effect}, we show the dependence of TAM accumulation in the $y$-direction on the chemical potential
for different angular momenta $j\in\{1/2,3/2,5/2\}$. Notably, the
Edelstein effect is enhanced and exhibits a broader plateau when the
TAM degrees of freedom are enlarged and the states are mostly occupied,
as seen by comparing the blue and black lines in Fig.~\ref{fig:Edelstein_effect}. 

There is no Edelstein effect for $j=1/2$ states in the energy window
$\mu_{c}/E_{F}\in[0,0.7]$, see red line in Fig.~\ref{fig:Edelstein_effect}(a). In this regime, the Fermi level lies below the onset of the Rashba-split bands so that no Fermi contour contributes to the response. As the chemical
potential increases, the electric field modifies the occupation of
states near the Fermi level and a TAM accumulation forms along the
$y$-direction. For $\mu_{c}/E_{F}\gtrsim0.7$, only the lower branch
$E_{-,\mathbf{k}_{\perp}}$, which is partially occupied, intersects
the Fermi level and contributes to $\chi_{xy}^{(1/2)}$. In this case,
the two Fermi momenta (which we denote by $|\mathbf{k}_{F}^{\pm}|$
for the inner and outer radii) originate from a $\emph{single}$ energy
band whose minimum occurs at finite $|\mathbf{k}_{\perp}|$, i.e.,
a Mexican-hat-like dispersion. Consequently, $\chi_{xy}^{(1/2)}$
depends strongly on the chemical potential and grows with $\mu_{c}$
until the upper branch $E_{+,\mathbf{k}_{\perp}}$ also becomes occupied.

Once both helical branches are present at the Fermi level, the difference
between the spin-split Fermi radii becomes independent of $\mu_{c}$,
and the Edelstein response saturates at a plateau. This is because the Rashba spin splitting is isotropic in momentum space \citep{isotropic}. This can be analytically deduced from the spin-$1/2$ model Hamiltonian that includes both the kinetic and Rashba SOC, i.e.,
\begin{equation}
H_{0}=\mathcal{A}_{1}+\gamma_{1,1}(k_{x}\hat{J}_{y}-k_{y}\hat{J}_{x}),
\end{equation}
where $\mathcal{A}_{1}=\mu_{1}+a_{1,1}|\mathbf{k}_{\perp}|^{2}$ and
$\hat{J}_{i}\rightarrow\hat{\sigma}_i$ is the Pauli matrices. The spectrum for $H_{0}$ is
$E_{n,\mathbf{k}_{\perp}}=\mathcal{A}_{1}+n\gamma_{1,1}|\mathbf{k}_{\perp}|$
where $|\mathbf{k}_{\perp}|=\sqrt{k_{x}^{2}+k_{y}^{2}}$ and $n=\pm$.
The relevant group velocity vector is $\mathbf{v}_{n}=(v_{n}^{(x)},v_{n}^{(y)})$
and $v_{n}^{(i)} = 2a_{1,1}k_{i} + n\gamma_{1,1} {k_{i}}/{|\mathbf{k}_{\perp}|}$
with $i\in\{x,y\}$. In addition, the TAM texture vector for $H_{0}$
becomes $\mathbf{J}_{\pm} = \pm ({1}/{|\mathbf{k}_{\perp}|})(-k_{y},k_{x})$
that is perpendicular to the momentum direction, i.e., $\mathbf{J}_{\pm}\cdot\mathbf{k}_{\perp}=0$.
The Fermi contours are circular $|\mathbf{k}_{F}^{\pm}|^{2}\equiv k_{x}^{2}+k_{y}^{2}$
with radius 
\begin{align}
|\mathbf{k}_{n,F}^{(l)}| &= \frac{1}{2a_{1,1}}
  \left(\! -n\gamma_{1,1} + (-1)^{l}\sqrt{1 + 4a_{1,1}(\mu_{c} -\mu_{1})} \right),
\end{align}
where $l \in \{1,2\}$ denoting two root solutions per given band. Considering
polar coordinate and after some algebra, we obtain the Edelstein susceptibility
\begin{equation}
  \chi_{xy}^{(1/2)}/\pi=-\sum_{n}n\,\text{sgn}(2a_{1,1}|\mathbf{k}_{F}^{n}|+n\gamma_{1,1})|\mathbf{k}_{F}^{n}|, 
\end{equation}
where it results in 
\begin{equation}
\chi_{xy}^{(1/2)}/\mathrm{c}_{0}\pi=|\mathbf{k}_{F}^{-}|-|\mathbf{k}_{F}^{+}|=\frac{\gamma_{1,1}}{a_{1,1}},
\end{equation}
where $\chi_{xy}^{(1/2)}$ is independent of $\mu_{c}$ for $\mu_{c}>\mu_{1}$,
see the red plateau in Fig.~\ref{fig:Edelstein_effect}(a). At $\gamma_{1,1}=0$,
the response is vanishing $\chi_{xy}^{(1/2)}=0$ because $|\mathbf{k}_{F}^{-}|=|\mathbf{k}_{F}^{+}|$. 

Importantly, in $j>1/2$ systems this saturation is no longer perfectly
flat: $\chi_{xy}^{(j)}$ continues to vary smoothly with the chemical
potential, as seen from the black and blue lines in Fig.~\ref{fig:Edelstein_effect}(a).
This is because higher TAM introduces additional $m_{j}$-dependent
subbands and increases the number of Fermi contours, leading to a
more complex evolution of the Fermi surface as $\mu_{c}$ is tuned.

Importantly, when the multipolar SOC terms are included in the Hamiltonian, the Edelstein effect can be further enhanced for the
$j=3/2$ and $j=5/2$ electrons, see Fig.~\ref{fig:Edelstein_effect}(b). To see this, consider the multipolar SOC Hamiltonian 
\begin{equation}
H=H_{0}+\mathcal{A}_{2}\hat{J}_{z}^{2}+\gamma_{1,2}(k_{x}\lceil\hat{J}_{y}\hat{J}_{z}^{2}\rfloor-k_{y}\lceil\hat{J}_{x}\hat{J}_{z}^{2}\rfloor),
\end{equation}
where the second term represents $m_{j}$-dependent energy shifts,
and the third term is the multipolar Rashba effect for $j \in \{3/2,5/2\}$.
This modification changes the density of states near the Fermi level.
Additionally, the number of Fermi contours can increase to four (six)
through energy shifts $\mathcal{A}_{2}\hat{J}_{z}^{2}$, resulting
in larger values, i.e., $\langle J_{y}^{(5/2)}\rangle>\langle J_{y}^{(3/2)}\rangle>\langle J_{y}^{(1/2)}\rangle$.
In the limit $\gamma_{1,2}=0$, the enhancement of the Edelstein effect originates from the energy shifts for electrons with different $m_{j}$, induced by the multipolar
term $\hat{J}_{z}^{2}$. 

Note that the Edelstein effect is absent
in $\hat{\mathcal{H}}_{\text{HS}}^{(2)}$, since the SOC contribution
from the multipolar doublet generator $({1}/{2}\, \llbracket(\hat{J}_{y}^{2}-\hat{J}_{x}^{2})\hat{J}_{z}\rrbracket,\, \llbracket\hat{J}_{x}\hat{J}_{y}\hat{J}_{z}\rrbracket)$
is purely of interband character, with a vanishing in-plane TAM texture. Although
this doublet exhibits an out-of-plane TAM texture, an in-plane electric
field $E_x$ cannot generate an Edelstein response in the
$z$-direction because of the mirror symmetry in the $xz$-plane.
The linear response must be invariant under this mirror reflection so that $\langle J_{z}^{^{\prime}(j)}\rangle=\langle J_{z}^{(j)}\rangle$,
while the axial nature of TAM implies $\langle J_{z}^{^{\prime}(j)}\rangle=-\langle J_{z}^{(j)}\rangle$.
These two conditions are only compatible if $\chi_{zx}=0$.

\section{\label{sec:Conclusion}Conclusion}
In this work, we develop a theoretical framework to quantify spin-orbit
coupling in \textit{p}- and \textit{d}-electrons of heavy elements. We focus on
Bloch states constrained by time-reversal and $C_{3v}$ point-group
symmetry, as realized, for example, in $\text{PtBi}_{2}$ and $\text{BiTeI}$
\citep{Ishizaka2011,PtBi2_2019,PtBi2_2021,PtBi2_2024}. Within this
setting, the total angular momentum is conserved and the bands decompose
into well-defined $j$-multiplets such as $j=3/2$ and $j=5/2$. Our
multipolar SOC framework is particularly relevant when a conventional
Rashba description cannot fully capture the observed band splittings,
as realized in $\text{BiTeI}$ \citep{Multipolar_signature}.

We show that in strongly spin-orbit-coupled systems the TAM vector
forms a texture in reciprocal space, as an analogue of the conventional
spin-$1/2$ texture. When multipolar corrections dominate the dipolar
SOC energy, the TAM texture becomes anisotropic per energy bands.
In particular, the TAM texture of heavy-mass electrons differs from
that of light-mass electrons, in contrast to the dipolar Rashba
model. They exhibit different varieties of winding at small and large
momenta in the vicinity of the 3D $\Gamma$ point. In addition, we
quantify the substantial enhancement of spin-orbit splitting for electrons
with $j>1/2$. We further demonstrated that these multipolar energy
corrections can enhance spintronic responses, such as current-induced
polarization.

It is worth mentioning that resolving TAM multiplets experimentally
is challenging. X-ray magnetic circular dichroism is a powerful probe
of SOC in $j$-shell electrons, providing information on spin and
orbital moments through polarization-dependent absorption \citep{Thole1992PRL,Carra1993PRL}.
Resonant inelastic x-ray scattering at appropriate absorption edges
is complementary, giving bulk-sensitive access to intra-multiplet
and spin-orbit excitations characteristic of $j$-manifolds in correlated
$4d$/$5d$ or $4f$/$5f$ systems \citep{Kotani2001RMP,Ament2011RMP}.
We additionally propose the development of $j$-resolved scanning
probes with quadrupolar or more general multipolar magnetic configurations, for example based on $d$-wave altermagnets \citep{Mulipolar_tipp_2024},
tunable ferromagnetic alloys \citep{Mulipolar_tipp_2012}, or multipolar
complex oxides \citep{Mulipolar_tipp_2004}. In addition, the orbital component of $\langle\mathbf{J}\rangle$ in $j>1/2$ electrons is dominant, highlighting an intrinsically orbital-driven character of the TAM response and supporting orbitronic applications \citep{Orbit_2008,Orbit_2008_1,Orbit_2021,Orbit_2022,Orbit_2023,Orbit_2024,Orbit_2025}.
\section{Acknowledgments}
We thank Chang-An Li for fruitful discussions. This work was supported
by the Deutsche Forschungsgemeinschaft (DFG, German Research Foundation)
through SFB 1170 (project ID 258499086), 
SFB 1143 (project ID 247310070, project A04),
and the W{\"u}rzburg-Dresden Cluster
of Excellence ctd.qmat (EXC 2147, project ID 390858490).
\appendix
\section{\label{Appnd3}Decomposition of $j$ into $C_{3v}$ irreducible representations}
We assume that SOC is strong compared to the kinetic energy. In this
case, the TAM is conserved with values $J\in[|l-S|,|l+S|]$. The splitting
of an energy level, labeled with integer $J$, is the same as for
$l$. However, half-integer $J$ is not and it should be treated
using a double-group representation. In this case, the double group
character table for the $C_{3v}$ point group is given in Table~\ref{Table:Double_character}.
The number of irreps in this table is twice that of the ordinary (single)
character table. This doubling arises because a fermionic state does
not return to itself under a $\theta\rightarrow\theta+2\pi$ rotation;
instead, it requires a $\theta\rightarrow\theta+4\pi$ rotation to
remain invariant. In this context, the spin characters under a $2\pi$
rotation transform as
\begin{equation}
X^{J}(R_{\theta+2\pi})=(-1)^{2J}X^{J}(R_{\theta}),
\end{equation}
where $R_{\theta}\in\{E,\,2C_{3z},\,3\sigma_{v}\}$ and
\begin{equation}
X^{J}(R_{\theta})=\frac{\sin\left((J+1/2)\theta\right)}{\sin(\theta/2)}.
\end{equation}

To determine how the representation $D_{J}$ decomposes into the irreps
$\Gamma_{i}$, we use the decomposition formula $D_{J}=\sum_{i=1}^{6}a_{i}\Gamma_{i}$
where the coefficients are given by
\begin{equation}
a_{i}=\frac{1}{h}\sum N_{R_{\theta}}D_{J}^{*}(R_{\theta})\Gamma_{i}(R_{\theta}),\label{coefic_decompos}
\end{equation}
with $N_{R_{\theta}}$ denoting the number of occurrences of the symmetry
operation $R_{\theta}$ and $h=12$ being the order of the double
group. After plugging the results of Tables~\ref{Table:Double_character}
and \ref{Table:Double_character_J} into Eq.~$\text{(\ref{coefic_decompos})}$,
we obtain the decomposition of quantum states under strong SOC
\begin{align}
D_{1/2} & \rightarrow\Gamma_{4},\\
D_{3/2} & \rightarrow\Gamma_{4}+\Gamma_{5}+\Gamma_{6},\\
D_{5/2} & \rightarrow2\Gamma_{4}+\Gamma_{5}+\Gamma_{6},
\end{align}
where the $D_{1/2}$ states remain doubly degenerate labeled with
$\Gamma_{4}$ in the presence of a crystal field, while $D_{3/2}$
and $D_{5/2}$ states split into one (two) doubly degenerate representations
and two one-dimensional states corresponding to complex conjugate
irreps $\Gamma_{5}$ and $\Gamma_{6}$. These two irreps form a reducible
2D representation as $\Gamma_{3/2}\equiv\Gamma_{5}\oplus\Gamma_{6}$.
Note that $D_{3/2}$ ($D_{5/2}$) is a four-dimensional (six-dimensional)
representation of the
full rotational group comprising of four (six) magnetic sub-levels labeled
by $m_{j}\in\{\pm1/2,\pm3/2\}$ ($m_{j}\in\{\pm1/2,\pm3/2,\pm5/2\}$).
Importantly, the dimension of $\Gamma_{5}$ and $\Gamma_{6}$ are
restricted to $1$ because of the dimensionality $\sum_{i=1}^{6}l_{i}^{2}=h$
where $l_{i}$ is the dimension of irreps given in the second column
of Table~\ref{Table:Double_character}. Consequently, we
have $1^{2}+1^{2}+2^{2}+2^{2}+l_{5}^{2}+l_{6}^{2}=12$, resulting in $l_{5}^{2}+l_{6}^{2}=2$,
where $l_{5}=l_{6}=1$ is the only possible solution. 
\begin{table}
\caption{\label{Table:Double_character}Double-group character table of the point group $C_{3v}$
for treating half-integer quantum numbers at high
symmetry points $\Gamma$, $A$, $H$, $K$.}
\centering{}%
\begin{tabular}{ccccccc}
\hline 
\hline 
 & $E$ & $\bar{E}$ & $2C_{3z}$ & $2\bar{C}_{3z}$ & $3\sigma_{v}$ & $3\bar{\sigma}_{v}$\tabularnewline
\hline 
$\Gamma_{1}$ & $1$ & $1$ & $1$ & $1$ & $1$ & $1$\tabularnewline
\hline 
$\Gamma_{2}$ & $1$ & $1$ & $1$ & $1$ & $-1$ & $-1$\tabularnewline
\hline 
$\Gamma_{3}$ & $2$ & $2$ & $-1$ & $-1$ & $0$ & $0$\tabularnewline
\hline 
$\Gamma_{4}$ & $2$ & $-2$ & $1$ & $-1$ & $0$ & $0$\tabularnewline
\hline 
$\Gamma_{5}$ & $1$ & $-1$ & $-1$ & $1$ & $i$ & $-i$\tabularnewline
\hline 
$\Gamma_{6}$ & $1$ & $-1$ & $-1$ & $1$ & $-i$ & $i$\tabularnewline
\hline 
\hline 
\end{tabular}
\end{table}
\begin{table}[t]
\caption{\label{Table:Double_character_J}Double-group character table for half-integer total angular momentum $J\in\{1/2,3/2,5/2\}$.}
\centering{}%
\begin{tabular}{ccccccc}
\hline 
\hline 
 & $E$ & $\bar{E}$ & $2C_{3z}$ & $2\bar{C}_{3z}$ & $3\sigma_{v}$ & $3\bar{\sigma}_{v}$\tabularnewline
\hline  
$D_{1/2}$ & $2$ & $-2$ & $1$ & $-1$ & $0$ & $0$\tabularnewline
\hline 
$D_{3/2}$ & $4$ & $-4$ & $-1$ & $1$ & $0$ & $0$\tabularnewline
\hline 
$D_{5/2}$ & $6$ & $-6$ & $0$ & $0$ & $0$ & $0$\tabularnewline
\hline 
\hline 
\end{tabular}
\end{table}
\section{\label{Appnd1}Point-Group and Time-Reversal Constraints in the $j>1/2$ basis}

In this Appendix, we present a detailed construction of the $\mathbf{k} \cdot \mathbf{p}$
model that preserves both the $C_{3v}$ point-group symmetry and time-reversal
symmetry $T$. Using group-theoretical methods, we identify all momentum-dependent
basis polynomials and TAM-tensor matrices that transform under
the gray magnetic point group $\mathcal{M}_{3v}=C_{3v}+TC_{3v}$. The relevant character table is given in Table~\ref{Table:character}. The characters of magnetic irreps that are odd under time-reversal are obtained by flipping the sign of the characters of the corresponding time-reversal-even representations.
 By taking
direct products of these momenta and TAM representations, we
obtain the multipolar SOC terms near the $\Gamma$ point of the three-dimensional bulk Brillouin zone.

\begin{table}
\caption{\label{Table:character}Character table of $\mathcal{M}_{3v}$
gray magnetic point group. irreps are labeled
by their behavior under time-reversal symmetry, with the last column
indicating whether each representation is even or odd under the action
of $T$.}
\centering{}%
\begin{tabular}{ccccccc}
\hline 
\hline 
 & $E$ & $2C_{3z}$ & $3\sigma_{v}$ & $2TC_{3z}$ & $3T\sigma_{v}$ & $T$\tabularnewline
\hline 
$A_{1,+}$ & $+1$ & $+1$ & $+1$ & $+1$ & $+1$ & $+1$\tabularnewline
\hline 
$A_{2,+}$ & $+1$ & $+1$ & $-1$ & $+1$ & $-1$ & $+1$\tabularnewline
\hline 
$E_{+}$ & $+2$ & $-1$ & $0$ & $-1$ & $0$ & $+2$\tabularnewline
\hline 
$A_{1,-}$ & $+1$ & $+1$ & $+1$ & $-1$ & $-1$ & $-1$\tabularnewline
\hline 
$A_{2,-}$ & $+1$ & $+1$ & $-1$ & $-1$ & $+1$ & $-1$\tabularnewline
\hline 
$E_{-}$ & $+2$ & $-1$ & $0$ & $+1$ & $0$ & $-2$\tabularnewline
\hline 
\hline 
\end{tabular}
\end{table}
\begin{table*}
\caption{Wave vector $\mathbf{k}=(k_{x},k_{y},k_{z})$ and total-angular-momentum
vector $\hat{\mathbf{J}}=(\hat{J}_{x},\hat{J}_{y},\hat{J}_{z})$ under $\mathcal{M}_{3v}$ magnetic point group symmetry.}\label{table:jprime}
\centering{}%
\begin{tabular}{ccccccc}
\hline 
\hline 
\foreignlanguage{american}{} & \multicolumn{3}{c}{$\mathbf{k}$} & \multicolumn{3}{c}{$\hat{\mathbf{J}}$}\tabularnewline
\hline 
$E$ & $k_{x}$ & $k_{y}$ & $k_{z}$ & $\hat{J}_{x}$ & $\hat{J}_{y}$ & $\hat{J}_{z}$\tabularnewline
$C_{3z}$ & $\frac{-1}{2}k_{x}\!-\!\frac{\sqrt{3}}{2}k_{y}$ & $\frac{\sqrt{3}}{2}k_{x}\!-\!\frac{1}{2}k_{y}$ & $k_{z}$ & $\frac{-1}{2}\hat{J}_{x}\!+\!\frac{\sqrt{3}}{2}\hat{J}_{y}$ & $\frac{-\sqrt{3}}{2}\hat{J}_{x}\!-\!\frac{1}{2}\hat{J}_{y}$ & $\hat{J}_{z}$\tabularnewline
$C_{3z}^{2}$ & $\frac{-1}{2}k_{x}\!+\!\frac{\sqrt{3}}{2}k_{y}$ & $\frac{-\sqrt{3}}{2}k_{x}\!-\!\frac{1}{2}k_{y}$ & $k_{z}$ & $\frac{-1}{2}\hat{J}_{x}\!-\!\frac{\sqrt{3}}{2}\hat{J}_{y}$ & $\frac{\sqrt{3}}{2}\hat{J}_{x}\!-\!\frac{1}{2}\hat{J}_{y}$ & $\hat{J}_{z}$\tabularnewline
\hline  
$\sigma_{v_{1}}$ & $k_{x}$ & $-k_{y}$ & $k_{z}$ & $-\hat{J}_{x}$ & $\hat{J}_{y}$ & $-\hat{J}_{z}$\tabularnewline
$\sigma_{v_{2}}$ & $\frac{-1}{2}k_{x}\!+\!\frac{\sqrt{3}}{2}k_{y}$ & $\frac{\sqrt{3}}{2}k_{x}\!+\!\frac{1}{2}k_{y}$ & $k_{z}$ & $\frac{1}{2}\hat{J}_{x}\!-\!\frac{\sqrt{3}}{2}\hat{J}_{y}$ & $\frac{-\sqrt{3}}{2}\hat{J}_{x}\!-\!\frac{1}{2}\hat{J}_{y}$ & $-\hat{J}_{z}$\tabularnewline
$\sigma_{v_{3}}$ & $\frac{-1}{2}k_{x}\!-\!\frac{\sqrt{3}}{2}k_{y}$ & $\frac{-\sqrt{3}}{2}k_{x}\!+\!\frac{1}{2}k_{y}$ & $k_{z}$ & $\frac{1}{2}\hat{J}_{x}\!+\!\frac{\sqrt{3}}{2}\hat{J}_{y}$ & $\frac{\sqrt{3}}{2}\hat{J}_{x}\!-\!\frac{1}{2}\hat{J}_{y}$ & $-\hat{J}_{z}$\tabularnewline
\hline 
$TC_{3z}$ & $\frac{1}{2}k_{x}\!+\!\frac{\sqrt{3}}{2}k_{y}$ & $\frac{1}{2}k_{y}\!-\!\frac{\sqrt{3}}{2}k_{x}$ & $-k_{z}$ & $\frac{1}{2}\hat{J}_{x}\!-\!\frac{\sqrt{3}}{2}\hat{J}_{y}$ & $\frac{\sqrt{3}}{2}\hat{J}_{x}\!+\!\frac{1}{2}\hat{J}_{y}$ & $-\hat{J}_{z}$\tabularnewline
$TC_{3z}^{2}$ & $\frac{1}{2}k_{x}\!-\!\frac{\sqrt{3}}{2}k_{y}$ & $\frac{\sqrt{3}}{2}k_{x}\!+\!\frac{1}{2}k_{y}$ & $-k_{z}$ & $\frac{1}{2}\hat{J}_{x}\!+\!\frac{\sqrt{3}}{2}\hat{J}_{y}$ & $\frac{-\sqrt{3}}{2}\hat{J}_{x}\!+\!\frac{1}{2}\hat{J}_{y}$ & $-\hat{J}_{z}$\tabularnewline
\hline 
$T\sigma_{v_{1}}$ & $-k_{x}$ & $k_{y}$ & $-k_{z}$ & $\hat{J}_{x}$ & $-\hat{J}_{y}$ & $\hat{J}_{z}$\tabularnewline
$T\sigma_{v_{2}}$ & $\frac{1}{2}k_{x}\!-\!\frac{\sqrt{3}}{2}k_{y}$ & $\frac{-\sqrt{3}}{2}k_{x}\!-\!\frac{1}{2}k_{y}$ & $-k_{z}$ & $\frac{-1}{2}\hat{J}_{x}\!+\!\frac{\sqrt{3}}{2}\hat{J}_{y}$ & $\frac{\sqrt{3}}{2}\hat{J}_{x}\!+\!\frac{1}{2}\hat{J}_{y}$ & $\hat{J}_{z}$\tabularnewline
$T\sigma_{v_{3}}$ & $\frac{1}{2}k_{x}\!+\!\frac{\sqrt{3}}{2}k_{y}$ & $\frac{\sqrt{3}}{2}k_{x}\!-\!\frac{1}{2}k_{y}$ & $-k_{z}$ & $\frac{-1}{2}\hat{J}_{x}\!-\!\frac{\sqrt{3}}{2}\hat{J}_{y}$ & $\frac{-\sqrt{3}}{2}\hat{J}_{x}\!+\!\frac{1}{2}\hat{J}_{y}$ & $\hat{J}_{z}$\tabularnewline
\hline 
$T$ & $-k_{x}$ & $-k_{y}$ & $-k_{z}$ & $-\hat{J}_{x}$ & $-\hat{J}_{y}$ & $-\hat{J}_{z}$\tabularnewline
\hline 
\hline 
\end{tabular}
\end{table*}

We focus on TAM values $j\in\{{3}/{2},{5}/{2}\}$. This
choice is motivated by the strong interaction between the electron spin and the
$d$-orbital degrees of freedom. Our goal is to construct the analogue
of the Pauli spin vector in a $j>1/2$ basis, i.e.,
\begin{equation}
\hat{\mathbf{J}}=(\hat{J}_{x},\hat{J}_{y},\hat{J}_{z}),
\end{equation}
with $[\hat{J}_{i},\hat{J}_{j}]=i\epsilon_{ijk}\hat{J}_{k}.$ The
eigenstates $\lvert j,m_{j}\rangle$ satisfy
\begin{equation}
\hat{J}_{z}\lvert j,m_{j}\rangle=m_{j}\lvert j,m_{j}\rangle
\end{equation}
so that $\hat{J}_{z}$ is diagonal in this basis,
\begin{equation}
\hat{J}_{z}=\frac{1}{2}\,\text{diag}(5,3,1,-1,-3,-5).
\label{Jz52}
\end{equation}
To obtain the $x$- and $y$-components of the vector operator $\hat{\mathbf{J}}$,
we introduce the ladder operators
\begin{equation}
\hat{J}_{\pm}=\hat{J}_{x}\pm i\hat{J}_{y},
\end{equation}
which act on the basis states as
\begin{equation}
J_{\pm}|j,m_{j}\rangle = \sqrt{j(j+1)-m_{j}(m_{j}\pm1)}\:
  |j,m_{j}\pm1\rangle,\label{eq:LadderEigenvalue}
\end{equation}
where $[J_{z},J_{\pm}]=\pm J_{\pm}$. Using Eq.~$\text{(\ref{eq:LadderEigenvalue})}$,
we obtain matrix form for $\hat{J}_{x(y)}$, for instance in $j=5/2$
basis, as given by
\begin{align}
\hat{J}_{x} &= \left(\!\!\begin{array}{cccccc}
\!0 & \!\frac{\sqrt{5}}{2} & \!0 & \!0 & \!0 & \!0\\
\!\frac{\sqrt{5}}{2} & \!0 & \!\sqrt{2} & \!0 & \!0 & \!0\\
\!0 & \!\sqrt{2} & \!0 & \!\frac{3}{2} & \!0 & \!0\\
\!0 & \!0 & \!\frac{3}{2} & \!0 & \!\sqrt{2} & \!0\\
\!0 & \!0 & \!0 & \!\sqrt{2} & \!0 & \!\frac{\sqrt{5}}{2}\\
\!0 & \!0 & \!0 & \!0 & \!\frac{\sqrt{5}}{2} & \!0
\end{array}\!\!\right) ,\label{Jx52} \\
\hat{J}_{y} &= \left(\!\!\begin{array}{cccccc}
\!0 & \!\frac{i\sqrt{5}}{2} & \!0 & \!0 & \!0 & \!0\\
\!\frac{-i\sqrt{5}}{2} & \!0 & \!i\sqrt{2} & \!0 & \!0 & \!0\\
\!0 & \!-i\sqrt{2} & \!0 & \!\frac{3i}{2} & \!0 & \!0\\
\!0 & \!0 & \!\frac{-3i}{2} & \!0 & \!\!i\sqrt{2} & \!0\\
\!0 & \!0 & \!0 & \!-i\sqrt{2} & \!0 & \!\frac{i\sqrt{5}}{2}\\
\!0 & \!0 & \!0 & \!0 & \!\frac{-i\sqrt{5}}{2} & \!0
\end{array}\!\!\right) .
\label{Jy52}
\end{align}
To analyze the symmetry operations of the magnetic point group $\mathcal{M}_{3v}$,
we express its elements as matrices in the $j>1/2$ basis. The nonmagnetic
subgroup $C_{3v}$ consists of
\begin{equation}
g\in\{E,C_{3z},C_{3z}^{2},\sigma_{v_{1}},\sigma_{v_{2}},\sigma_{v_{3}}\},\label{eq:GroupC3velements}
\end{equation}
where $E$ is the identity operation and $C_{3z}$ ($C_{3z}^{2}$)
denotes a threefold rotation around the $z$-axis by an angle $\theta=2\pi/3$ ($4\pi/3$).
The matrix representation of $C_{3z}$ is given by
\begin{equation}
\hat{C}_{3z} \equiv e^{-2\pi i\hat{J}_{z}/3} = \left(\begin{array}{cccccc}
\nu & 0 & 0 & 0 & 0 & 0\\
0 & -1 & 0 & 0 & 0 & 0\\
0 & 0 & \nu^{*} & 0 & 0 & 0\\
0 & 0 & 0 & \nu & 0 & 0\\
0 & 0 & 0 & 0 & -1 & 0\\
0 & 0 & 0 & 0 & 0 & \nu^{*}
\end{array}\right),
\end{equation}
where $\nu=e^{{i\pi}/{3}}$. In Eq.~(\ref{eq:GroupC3velements}),
the mirror reflection $\sigma_{v_{1}}$ is implemented in the TAM basis as a $\pi$-rotation about the $y$-axis, i.e.,
\begin{equation}
\hat{\sigma}_{v_{1}} \equiv e^{-i\pi\hat{J}_{y}} = \left(\begin{array}{cccccc}
0 & 0 & 0 & 0 & 0 & 1\\
0 & 0 & 0 & 0 & -1 & 0\\
0 & 0 & 0 & 1 & 0 & 0\\
0 & 0 & -1 & 0 & 0 & 0\\
0 & 1 & 0 & 0 & 0 & 0\\
-1 & 0 & 0 & 0 & 0 & 0
\end{array}\right) .
\end{equation}
This implies that under the $\sigma_{v_{1}}$ operation, the wave
vector and the TAM vector transform differently.
\begin{align}
\sigma_{v_1} &: \left\{
  \begin{array}{ll}
    \mathbf{k} &\rightarrow (+k_{x},-k_{y},+k_{z}), \\[0.5ex]
    \hat{\mathbf{J}} &\rightarrow (-\hat{J}_{x},\hat{J}_{y},-\hat{J}_{z}).
  \end{array}\right.
\end{align}
The other two mirror operators, $\hat{\sigma}_{v_{2}}$ and $\hat{\sigma}_{v_{3}}$,
can be obtained by rotating $\hat{\sigma}_{v_{1}}$ around the $z$-axis
by an angle of $\theta=2\pi/3$ and $\theta=4\pi/3$, respectively
given by
\begin{align}
\!\!\!\hat{\sigma}_{v_{2}} & \!=\!\hat{C}_{3z}\hat{\sigma}_{v_{1}}[\hat{C}_{3z}]^{\dagger}\!=\!\left(\!\!\begin{array}{cccccc}
0 & 0 & 0 & 0 & 0 & \nu^{\prime}\\
0 & 0 & 0 & 0 & -1 & 0\\
0 & 0 & 0 & \nu^{\prime}{}^{*} & 0 & 0\\
0 & 0 & -\nu^{\prime} & 0 & 0 & 0\\
0 & 1 & 0 & 0 & 0 & 0\\
-\nu^{\prime}{}^{*} & 0 & 0 & 0 & 0 & 0
\end{array}\!\!\right)\!,\!\!\\
\!\!\!\hat{\sigma}_{v_{3}}\! & =\!\hat{C}_{3z}^{2}\hat{\sigma}_{v_{1}}[\hat{C}_{3z}^{2}]^{\dagger}\!=\!\hat{\sigma}_{v_{2}}^{*},\!\!
\end{align}
where $\nu^{\prime}=e^{2i\pi/3}$. To proceed further, it is important
to understand how the TAM     vector $\hat{\mathbf{J}}$ and the wave
vector $\mathbf{k}$ transform under the magnetic point group $\mathcal{M}_{3v}$.
In general, both vectors transform according to the symmetry operations
of $\mathcal{M}_{3v}$ such that
\begin{align}
\hat{g}\hat{\mathbf{J}}\hat{g}^{\dagger} &\mapsto \hat{\mathbf{J}}^{\prime},\\
\hat{R}^{-1}\mathbf{k} &\mapsto \mathbf{k}^{\prime}.
\end{align}
The transformation properties of $\hat{\mathbf{J}}^{\prime}$ and
$\mathbf{k}^{\prime}$ under all symmetry operations are summarized
in Table~\ref{table:jprime}. 

Importantly, $\hat{J}_{0}$ and $\hat{J}_{z}$ transform according
to the irreps $A_{1,+}$ and $A_{2,-}$,
respectively, while $(\hat{J}_{x},\hat{J}_{y})$ form a doublet transforming
under the $E_{-}$ irrep. The subscript of irreps, namely $E_-$ ($E_+$),
denotes the character of the time-reversal operator, which can be odd (even).
In addition, the momentum components $(k_{x},k_{y})$ transform as
the $E_{-}$ irrep and $k_{z}$ transforms according to the $A_{1,-}$
irrep. These transformation properties provide a key starting point
for constructing higher-order momentum-dependent polynomial terms
that are compatible with the symmetry of the magnetic group $\mathcal{M}_{3v}$.
The resulting polynomials, up to fifth order, are summarized in Table~\ref{table:M3V momentum dependent polynomials},
and are derived by applying direct products of irreps as detailed
in Table~\ref{DirectProductMagnetic}. For instance, a second-order polynomial that transforms according to the $E_{+}$ irrep can be constructed by taking the direct product
of two basis functions transforming as $E_{-}$. This relation is
expressed as $E_{+}=E_{-}\otimes E_{-}$, which which the relevant characters are listed in Table~\ref{tab:Eminus_tensor_Eminus},
\begin{table}[t]
\centering
\caption{Character of $E_{-}\otimes E_{-}$ product.}
\label{tab:Eminus_tensor_Eminus}
\begin{tabular}{ccccccc}
\hline\hline
 & $E$ & $C_{3z}$ & $\sigma_{v}$ & $TE$ & $TC_{3z}$ & $T\sigma_{v}$\\
\hline
$E_{-}$ & $2$ & $-1$ & $0$ & $-2$ & $1$ & $0$\\
\hline
$E_{-}\otimes E_{-}$ & $4$ & $1$ & $0$ & $4$ & $1$ & $0$\\
\hline\hline
\end{tabular}
\end{table}
where the first row lists the symmetry elements of the $\mathcal{M}_{3v}$
group, while the second and third rows show the character values
for the irrep $E_{-}$, and its direct product $E_{-}\otimes E_{-}$.
We infer that this direct product is a reducible representation,
as its dimensionality exceeds that of a typical two-dimensional irrep.
Specifically, the character of the identity element $E$ in Table~\ref{tab:Eminus_tensor_Eminus}
is four, indicating a four-dimensional representation. Therefore,
this reducible representation decomposes into a direct sum of irreps
as
\begin{equation}
E_{-}\otimes E_{-}\rightarrow A_{1,+}\oplus A_{2,+}\oplus E_{+}.
\end{equation}
This decomposition is confirmed by summing the characters of the corresponding
irreps, as listed in Table~\ref{tab:OplusDeco}.
\begin{table}[t]
\centering
\caption{Character for direct sum of irreps from decomposition $E_{+}\otimes E_{+}$.}
\label{tab:OplusDeco}
\centering
\begin{tabular}{ccccccc}
\hline
\hline
 & $E$ & $C_{3z}$ & $\sigma_{v}$ & $TE$ & $TC_{3z}$ & $T\sigma_{v}$\\
 \hline
$A_{1,+}$ & $1$ & $1$ & $1$ & $1$ & $1$ & $1$\\
\hline
$A_{2,+}$ & $1$ & $1$ & $-1$ & $1$ & $1$ & $-1$\\
\hline
$E_{+}$   & $2$ & $-1$ & $0$ & $2$ & $-1$ & $0$\\
\hline
 $A_{1,+}\oplus A_{2,+}\oplus E_{+}$          & $4$ & $1$  & $0$ & $4$ & $1$  & $0$\\
 \hline
 \hline
\end{tabular}
\end{table}
In accordance, the reducible representation $E_{+} =E_{-}\otimes E_{-}$ contains four
momentum-dependent components, expressed as
\begin{equation}
(k_{x},k_{y})\otimes(k_{x},k_{y}) =(k_{x}^{2},k_{y}^{2},k_{x}k_{y},k_{y}k_{x}).\label{eq:E+reducibleMomentum}
\end{equation}
Note that only certain linear combinations of these components transform
according to the $E_{+}$ irrep. In particular, the pair $(k_{x}^{2}-k_{y}^{2},\;k_{x}k_{y})$
transforms as the $E_{+}$ irrep, while the scalar $k_{x}^{2}+k_{y}^{2}$
transforms according to $A_{1,+}$. However, it is not possible to
form a second-order polynomial from these terms that transforms as
$A_{2,+}$. Therefore, the $A_{2,+}$ representation does not appear
at second order in momentum, see Table~\ref{table:M3V momentum dependent polynomials}. 
\begin{table}
\caption{Direct product of irreducible representations in $\mathcal{M}_{3v}$
gray magnetic group. Subscript follows product rules $1\otimes2=2$, $1\otimes1=1$, $2\otimes2=1$,  $\pm\otimes\pm=+$, and $-\otimes+=-$.
$A_{1,+}$ is shown in bold, indicating symmetry-allowed SOC terms
from the corresponding direct product. We have defined $Q_{-}\equiv A_{1,-}\oplus A_{2,-}\oplus E_{-}$
and $\mathbf{Q}_{+}\equiv\boldsymbol{A_{1,+}}\oplus A_{2,+}\oplus E_{+}$.}\label{DirectProductMagnetic}
\centering{}%
\begin{tabular}{ccccccc}
\hline 
\hline 
\foreignlanguage{american}{} & $A_{1,+}$ & $A_{2,+}$ & $E_{+}$ & $A_{1,-}$ & $A_{2,-}$ & $E_{-}$\tabularnewline
\hline 
$A_{1,+}$ & \textbf{$\boldsymbol{A_{1,+}}$} & $A_{2,+}$ & $E_{+}$ & $A_{1,-}$ & $A_{2,-}$ & $E_{-}$\tabularnewline
\hline 
$A_{2,+}$ & $A_{2,+}$ & \textbf{$\boldsymbol{A_{1,+}}$} & $E_{+}$ & $A_{2,-}$ & $A_{1,-}$ & $E_{-}$\tabularnewline
\hline 
$E_{+}$ & $E_{+}$ & $E_{+}$ & \foreignlanguage{american}{$\mathbf{Q}_{+}$} & $E_{-}$ & $E_{-}$ & \foreignlanguage{american}{$Q_{-}$}\tabularnewline
\hline 
$A_{1,-}$ & $A_{1,-}$ & $A_{2,-}$ & $E_{-}$ & $\boldsymbol{A_{1,+}}$ & $A_{2,+}$ & $E_{+}$\tabularnewline
\hline 
$A_{2,-}$ & $A_{2,-}$ & $A_{1,-}$ & $E_{-}$ & $A_{2,+}$ & $\boldsymbol{A_{1,+}}$ & $E_{+}$\tabularnewline
\hline 
$E_{-}$ & $E_{-}$ & $E_{-}$ & \foreignlanguage{american}{$Q_{-}$} & $E_{+}$ & $E_{+}$ & \foreignlanguage{american}{$\mathbf{Q}_{+}$}\tabularnewline
\hline 
\hline 
\end{tabular}
\end{table}
\begin{table}
\caption{Momentum-dependent polynomials transforming according to $\mathcal{M}_{3v}$
magnetic point group. Order of each polynomial corresponds to order
of associated spherical harmonics. A cross symbol $(\times)$ indicates
that no polynomial basis exists for given irreducible representation.}\label{table:M3V momentum dependent polynomials}
\centering{}%
\begin{tabular}{cccc}
\hline 
\hline 
$A_{1,+}$ & $1$ & $k_{z}^{2}$ & $k_{z}^{4}$,$k_{z}(k_{x}^{3}\!-\!3k_{x}k_{y}^{2})$\tabularnewline
\hline 
$A_{2,+}$ & $\times$ & $\times$ & $k_{z}(3k_{y}k_{x}^{2}\!-\!k_{y}^{3})$\tabularnewline
\hline 
$E_{+}$ & $\times$ & $\Big(\!\begin{array}{c}
k_{x}^{2}\!-\!k_{y}^{2}\\
k_{x}k_{y}
\end{array}\!\Big)$ & $\Big(\!\begin{array}{c}
[k_{x}^{2}\!-\!k_{y}^{2}]k_{z}^{2}\\
k_{x}k_{y}k_{z}^{2}
\end{array}\!\Big)$\tabularnewline
\foreignlanguage{american}{} & $\times$ & $\Big(\!\begin{array}{c}
k_{x}k_{z}\\
k_{y}k_{z}
\end{array}\!\Big)$ & $\Big(\!\begin{array}{c}
k_{x}k_{z}^{3}\\
k_{y}k_{z}^{3}
\end{array}\!\Big)$\tabularnewline
\foreignlanguage{american}{} & $\times$ & $\times$ & $\Big(\!\begin{array}{c}
[k_{x}^{2}\!-\!k_{y}^{2}]^{2}\!-\!4k_{x}^{2}k_{y}^{2}\\
k_{x}k_{y}[k_{x}^{2}-k_{y}^{2}]
\end{array}\!\Big)$\tabularnewline
\hline 
$A_{1,-}$ & $k_{z}$ & $k_{x}^{3}-3k_{x}k_{y}^{2},$$k_{z}^{3}$ & $k_{z}^{2}(k_{x}^{3}\!-\!3k_{x}k_{y}^{2}),k_{z}^{5}$\tabularnewline
\hline 
$A_{2,-}$ & $\times$ & $3k_{y}k_{x}^{2}-k_{y}^{3}$ & $(3k_{y}k_{x}^{2}\!-\!k_{y}^{3})k_{z}^{2}$\tabularnewline
\hline 
$E_{-}$ & $\Big(\!\begin{array}{c}
k_{x}\\
k_{y}
\end{array}\!\Big)$ & $\Big(\!\begin{array}{c}
[k_{x}^{2}\!-\!k_{y}^{2}]k_{z}\\
k_{x}k_{y}k_{z}
\end{array}\!\Big)$ & $\Big(\!\begin{array}{c}
[k_{x}^{2}\!-\!k_{y}^{2}]k_{z}^{3}\\
k_{x}k_{y}k_{z}^{3}
\end{array}\!\Big)$\tabularnewline
\foreignlanguage{american}{} & $\times$ & $\Big(\!\begin{array}{c}
k_{x}k_{z}^{2}\\
k_{y}k_{z}^{2}
\end{array}\!\Big)$ & $\Big(\!\begin{array}{c}
k_{x}k_{z}^{4}\\
k_{y}k_{z}^{4}
\end{array}\!\Big)$\tabularnewline
\foreignlanguage{american}{} & $\times$ & $\times$ & $\Big(\!\begin{array}{c}
k_{x}^{5}\!-\!10k_{x}^{3}k_{y}^{2}\!+\!5k_{x}k_{y}^{4}\\
5k_{x}^{4}k_{y}\!-\!10k_{x}^{2}k_{y}^{3}+k_{y}^{5}
\end{array}\!\Big)$\tabularnewline
\foreignlanguage{american}{} & $\times$ & $\times$ & $\Big(\!\begin{array}{c}
[k_{x}^{2}\!-\!k_{y}^{2}]^{2}k_{z}\!-\!4k_{x}^{2}k_{y}^{2}k_{z}\\
k_{x}k_{y}k_{z}[k_{x}^{2}\!-\!k_{y}^{2}]
\end{array}\!\Big)$\tabularnewline
\hline 
\hline 
\end{tabular}
\end{table}

\section{\label{Appnd2}Multipolar SOC Construction in $\mathcal{M}_{3v}$ Symmetry}

To construct a 3D bulk model Hamiltonian, we utilize the momentum-dependent
polynomials listed in Table~\ref{table:M3V momentum dependent polynomials}
and combine them with the appropriate TAM tensor matrices transforming
according to the $A_{1,+}$ irrep. Since the SOC Hamiltonian must be Hermitian,
the Hermiticity is ensured by applying proper symmetrization procedures.
After a straightforward but lengthy algebraic process, the resulting
TAM tensor matrices are summarized in Table~\ref{tab:Spin-tensor-matrices}.

The relevant TAM basis functions are obtained by substituting momentum
components with TAM operators, i.e., $k_{i}\rightarrow\hat{J}_{i}$
for $i\in\{x,y,z\}$. To guarantee Hermitian structure, one must symmetrize
products of these operators appropriately. For example, both $k_{z}$
and the third-order polynomial $k_{x}^{3}-3k_{x}k_{y}^{2}$ transform
as the $A_{1,-}$ irrep, as indicated in Table~\ref{table:M3V momentum dependent polynomials}.
Interestingly, the corresponding TAM tensor matrices transform as
$A_{2,-}$ representations, which contrasts with the momentum basis
behavior, see Table~\ref{tab:Spin-tensor-matrices}. This is given
by
\begin{align}
A_{1-} \rightarrow A_{2-} &: \left\{
  \begin{array}{l}
    k_{z} \rightarrow \hat{J}_{z}, \\[0.5ex]
    k_{x}^{3}-3k_{x}k_{y}^{2} \rightarrow \llbracket\hat{J}_{x}^{3}-3\hat{J}_{x}\hat{J}_{y}^{2}\rrbracket,
  \end{array}\right. \label{eq:EplusKandJ}\\
E_+ \rightarrow E_+ &: \left\{
  \begin{array}{l}
    (k_{x}k_{z},k_{y}k_{z}) \rightarrow (\lceil\hat{J}_{x}\hat{J}_{z}\rfloor,\lceil\hat{J}_{y}\hat{J}_{z}\rfloor), \\[0.5ex]
    (k_{x}^{2}-k_{y}^{2},k_{x}k_{y}) \rightarrow (\hat{J}_{x}^{2}\!-\!\hat{J}_{y}^{2},\lceil\hat{J}_{x}\hat{J}_{y}\rfloor).
  \end{array}\right.
\end{align} 

The multipolar SOC model Hamiltonian is given by
\begin{equation}
H_{N}=\sum_{\mathbf{k}}\hat{\psi}_{\mathbf{k}}^{\dagger}\hat{H}(\mathbf{k})\hat{\psi}_{\mathbf{k}},
\end{equation}
where $\hat{\psi}_{\mathbf{k}}$ is a $6 \times 1$ ($4 \times 1$) spinor defined in the $j=5/2$ ($3/2$) TAM basis given by
\begin{align}
\hat{\psi}_{\mathbf{k}} &= (c_{\mathbf{k},{5}/{2}},c_{\mathbf{k},{3}/{2}},c_{\mathbf{k},{1}/{2}},c_{\mathbf{k},-{1}/{2}},c_{\mathbf{k},-{3}/{2}},c_{\mathbf{k},-{5}/{2}})^{T}, \\
\hat{\psi}_{\mathbf{k}} &= (c_{\mathbf{k},{3}/{2}},c_{\mathbf{k},{1}/{2}},c_{\mathbf{k},-{1}/{2}},c_{\mathbf{k},-{3}/{2}})^{T},
\end{align}
where $c_{\mathbf{k},m_{j}}^{\dagger}$ ($c_{\mathbf{k},m_{j}}$)
denotes the fermionic creation (annihilation) operator associated
with momentum $\mathbf{k}$ and magnetic quantum number $m_{j}\in[-j,j]$.
The Hamiltonian $\hat{H}(\mathbf{k})$ is a Hermitian matrix-valued
function and any term appearing in it must result from a combination
of momentum-dependent polynomials and TAM basis matrices that
transform according to the trivial irrep $A_{1,+}$,
satisfied by
\begin{equation}
\hat{g}\hat{H}(\hat{R}^{-1}\mathbf{k})\hat{g}^{\dagger} \mapsto +\hat{H}(\mathbf{k}),
\end{equation}
where $R^{-1}$ is the $3\times3$ orthogonal rotation matrix in momentum
space and $\hat{g}$ denotes the matrix form for
symmetry elements of $\mathcal{M}_{3v}$. 

Using Table~\ref{DirectProductMagnetic},
the multipolar SOC energy takes the form
\begin{align}
\hat{H}(\mathbf{k}) &= A_{1,+}^{m}\otimes A_{1,+}^{j}\!+\!A_{2,+}^{m}\otimes A_{2,+}^{j}\!+\!E_{+}^{m}\otimes E_{+}^{j}\nonumber \\
 &\quad{} +\!A_{1,-}^{m}\otimes A_{1,-}^{j}\!+\!A_{2,-}^{m}\otimes A_{2,-}^{j}\!+\!E_{-}^{m}\otimes E_{-}^{j}, \label{eq:NormalStateHamil}
\end{align}
where $A_{1,+}^{m(j)}$ denotes the symmetry-allowed momentum-dependent
polynomial (TAM tensor matrix) transforming according to the $A_{1,+}$
irrep, with the superscripts $m$ and $j$ labeling
momentum and TAM contributions, respectively. The first line in $\hat{H}(\mathbf{k})$
is even under parity exchange while the second line are odd in momentum.
These terms break inversion symmetry. After performing the necessary
symmetry analysis and algebraic construction, we arrive at the explicit
form of the even-parity SOC energy in the $j>1/2$ basis, given by
\begin{align}
A_{1,+}^{m}\otimes A_{1,+}^{j} &= \sum_{i=1}^{4}\mathbf{P}_{i}\cdot\mathbf{A}_{_{1+}}[\hat{\mathsf{A}}_{1+}]_{i}, \label{eq:A1pA1p}\\
A_{2,+}^{m}\otimes A_{2,+}^{j} &= k_{z}(3k_{y}k_{x}^{2}-k_{y}^{3})\lceil\llbracket3J_{y}J_{x}^{2} - J_{y}^{3}\rrbracket J_{z}\rfloor,\\
E_{+}^{m}\otimes E_{+}^{j} &= \sum_{i=1}^{4}\mathbf{N}_{i}\cdot(\mathbf{E}_{+}[\hat{\mathsf{E}}_{+}]_{i} +\mathbf{E}_{+}^{\prime}[\hat{\mathsf{E}}_{+}^{\prime}]_{i}).
\end{align}
Also, we have defined odd-parity terms in the normal state parts, which
are given by
\begin{align}
A_{1,-}^{m}\otimes A_{1,-}^{j} &= \sum_{i=1}^{2}\mathbf{T}_{i}\cdot\mathbf{A}_{_{1-}}[\hat{\mathsf{A}}_{1-}]_{i}, \\
A_{2,-}^{m}\otimes A_{2,-}^{j} &= \sum_{i=1}^{5}\mathbf{R}_{i}\cdot\mathbf{A}_{_{2-}}[\hat{\mathsf{A}}_{2-}]_{i}, \\
E_{-}^{m}\otimes E_{-}^{j} &= \sum_{i=1}^{5}\mathbf{G}_{i}\cdot(\mathbf{E}_{-}[\hat{\mathsf{E}}_{-}]_{i} + \mathbf{E}_{-}^{\prime}[\hat{\mathsf{E}}_{-}^{\prime}]_{i}),
\end{align}
\begin{table}
\caption{Total angular momentum tensor matrices for $j>1/2$ electrons transforming according to
$\mathcal{M}_{3v}$ point group.} \label{tab:Spin-tensor-matrices}
\centering{}%
\begin{tabular}{cccc}
\hline 
\hline 
$A_{1,+}$ & $\hat{J}_{0}$ & $\hat{J}_{z}^{2}$ & $\lceil\llbracket\hat{J}_{x}^{3}-3\hat{J}_{x}\hat{J}_{y}^{2}\rrbracket\hat{J}_{z}\rfloor$\tabularnewline
\foreignlanguage{american}{} & $\times$ & $\times$ & $\hat{J}_{z}^{4}$\tabularnewline
\hline 
$A_{2,+}$ & $\times$ & $\times$ & $\lceil\llbracket3\hat{J}_{y}\hat{J}_{x}^{2}-\hat{J}_{y}^{3}\rrbracket\hat{J}_{z}\rfloor$\tabularnewline
\hline 
$E_{+}$ & $\times$ & $\Big(\!\!\begin{array}{c}
\hat{J}_{x}^{2}-\hat{J}_{y}^{2}\\
\lceil\hat{J}_{x}\hat{J}_{y}\rfloor
\end{array}\!\!\Big)$ & $\Big(\!\!\begin{array}{c}
\lceil(\hat{J}_{x}^{2}-\hat{J}_{y}^{2})\hat{J}_{z}^{2}\rfloor\\
\lceil\lceil\hat{J}_{x}\hat{J}_{y}\rfloor\hat{J}_{z}^{2}\rfloor
\end{array}\!\!\Big)$\tabularnewline
\foreignlanguage{american}{} & $\times$ & $\Big(\!\!\begin{array}{c}
\lceil\hat{J}_{x}\hat{J}_{z}\rfloor\\
\lceil\hat{J}_{y}\hat{J}_{z}\rfloor
\end{array}\!\!\Big)$ & $\Big(\!\!\begin{array}{c}
\lceil\hat{J}_{x}\hat{J}_{z}^{3}\rfloor\\
\lceil\hat{J}_{y}\hat{J}_{z}^{3}\rfloor
\end{array}\!\!\Big)$\tabularnewline
\hline 
$A_{1,-}$ & $\times$ & $\llbracket3\hat{J}_{y}\hat{J}_{x}^{2}-\hat{J}_{y}^{3}\rrbracket$ & $\lceil\llbracket3\hat{J}_{y}\hat{J}_{x}^{2}-\hat{J}_{y}^{3}\rrbracket\hat{J}_{z}^{2}\rfloor$\tabularnewline
\hline 
$A_{2,-}$ & $\hat{J}_{z}$ & $\llbracket\hat{J}_{x}^{3}-3\hat{J}_{x}\hat{J}_{y}^{2}\rrbracket$ & $\lceil\llbracket\hat{J}_{x}^{3}-3\hat{J}_{x}\hat{J}_{y}^{2}\rrbracket\hat{J}_{z}^{2}\rfloor$\tabularnewline
\foreignlanguage{american}{} & $\times$ & $\hat{J}_{z}^{3}$ & $\hat{J}_{z}^{5}$\tabularnewline
\hline 
$E_{-}$ & $\Big(\!\!\begin{array}{c}
\hat{J}_{x}\\
\hat{J}_{y}
\end{array}\!\!\Big)$ & $\Big(\!\!\begin{array}{c}
\lceil(\hat{J}_{x}^{2}-\hat{J}_{y}^{2})\hat{J}_{z}\rfloor\\
\llbracket\hat{J}_{x}\hat{J}_{y}\hat{J}_{z}\rrbracket
\end{array}\!\!\Big)$ & $\Big(\!\!\begin{array}{c}
\lceil(J_{x}^{2}-J_{y}^{2})J_{z}^{3}\rfloor\\
\lceil\lceil J_{x}J_{y}\rfloor J_{z}^{3}\rfloor
\end{array}\!\!\Big)$\tabularnewline
\foreignlanguage{american}{} & $\times$ & $\Big(\!\!\begin{array}{c}
\lceil J_{x}J_{z}^{2}\rfloor\\
\lceil J_{y}J_{z}^{2}\rfloor
\end{array}\!\!\Big)$ & $\Big(\!\!\begin{array}{c}
\lceil J_{x}J_{z}^{4}\rfloor\\
\lceil J_{y}J_{z}^{4}\rfloor
\end{array}\!\!\Big)$\tabularnewline
\hline 
\hline 
\end{tabular}
\end{table}
Note that the full multipolar SOC Hamiltonian is a summation of odd-
and even-parity terms. The TAM basis elements that are even under
time-reversal symmetry are given by
\begin{align}
\hat{\mathsf{A}}_{1+} &\equiv (\hat{J}_{0},\hat{J}_{z}^{2},\hat{J}_{z}^{4}\lceil\llbracket\hat{J}_{x}^{3}-3\hat{J}_{x}\hat{J}_{y}^{2}\rrbracket\hat{J}_{z}\rfloor, \\
\hat{\mathsf{E}}_{+} &\equiv (\hat{J}_{x}^{2}-\hat{J}_{y}^{2},\lceil\hat{J}_{x}\hat{J}_{z}\rfloor,\lceil(\hat{J}_{x}^{2}-\hat{J}_{y}^{2})\hat{J}_{z}^{2}\rfloor,\lceil\hat{J}_{x}\hat{J}_{z}^{3}\rfloor), \\
\hat{\mathsf{E}}_{+}^{\prime} &\equiv (2\lceil\hat{J}_{x}\hat{J}_{y}\rfloor,-\lceil\hat{J}_{y}\hat{J}_{z}\rfloor,2\lceil\lceil\hat{J}_{x}\hat{J}_{y}\rfloor\hat{J}_{z}^{2}\rfloor,-\lceil\hat{J}_{y}\hat{J}_{z}^{3}\rfloor),
\end{align}
while the odd bases takes the form
\begin{align}
\!\!\!\hat{\mathsf{A}}_{1-}\!\equiv\!( & \llbracket3\hat{J}_{y}\hat{J}_{x}^{2}\!-\!\hat{J}_{y}^{3}\rrbracket,\lceil\llbracket3\hat{J}_{y}\hat{J}_{x}^{2}\!-\!\hat{J}_{y}^{3}\rrbracket\hat{J}_{z}^{2}\rfloor),\!\\
\!\!\!\hat{\mathsf{A}}_{2-}\!\equiv\!( & \hat{J}_{z},\llbracket\hat{J}_{x}^{3}\!-\!3\hat{J}_{x}\hat{J}_{y}^{2}\rrbracket,\hat{J}_{z}^{3},\lceil\llbracket\hat{J}_{x}^{3}\!-\!3\hat{J}_{x}\hat{J}_{y}^{2}\rrbracket\hat{J}_{z}^{2}\rfloor,\hat{J}_{z}^{5}),\!\\
\!\!\!\hat{\mathsf{E}}_{-}\!\equiv\!( & \hat{J}_{y},\llbracket\hat{J}_{x}\hat{J}_{y}\hat{J}_{z}\rrbracket,\lceil\hat{J}_{y}\hat{J}_{z}^{2}\rfloor,\lceil\lceil\hat{J}_{x}\hat{J}_{y}\rfloor\hat{J}_{z}^{3}\rfloor,\lceil\hat{J}_{y}\hat{J}_{z}^{4}\rfloor),\!\\
\!\!\!\hat{\mathsf{E}}_{-}^{\prime}\!\equiv\!( & \hat{J}_{x},-\frac{\lceil\hat{J}_{1}\hat{J}_{z}\rfloor}{2},\lceil\hat{J}_{x}\hat{J}_{z}^{2}\rfloor,-\frac{\lceil\hat{J}_{1}\hat{J}_{z}^{3}\rfloor}{2},\lceil\hat{J}_{x}\hat{J}_{z}^{4}\rfloor),\!\label{eq:EnegPrimeJJ}
\end{align}
where $\hat{J}_{1}\equiv\hat{J}_{x}^{2}-\hat{J}_{y}^{2}$.
Note that the coefficients of basis matrices in Eq.~$\text{(\ref{eq:EnegPrimeJJ})}$
are obtained to let Eq.~$\text{(\ref{eq:NormalStateHamil})}$ transform
according to $A_{1+}$ irrep. In addition, the relevant momentum-dependent
polynomials
are also compactly written in multi-element vectors defined by
\begin{align}
\!\!\!\!\!\mathbf{A}_{_{1+}}\! & \equiv\!(1,k_{x}^{2}\!+\!k_{y}^{2},k_{z}^{2},k_{z}^{4},k_{z}(k_{x}^{3}\!-\!3k_{x}k_{y}^{2})),\!\label{A1pluskk}\\
\!\!\!\!\!\mathbf{E}_{+}\! & \equiv\!(k_{x}k_{z},K_{1},K_{1}k_{z}^{2},[K_{1}^{2}-4K_{2}^{2}],k_{x}k_{z}^{3}),\!\\
\!\!\!\!\!\mathbf{E}_{+}^{\prime}\! & \equiv\!(-k_{y}k_{z},2K_{2},2K_{2}k_{z}^{2},-4K_{1}K_{2},-k_{y}k_{z}^{3}),\!
\end{align}
where $K_{1}\equiv k_{x}^{2}\!-\!k_{y}^{2}$ and $K_{2}\equiv k_{x}k_{y}$,
and the odd parity form factors become
\begin{align}
\!\!\!\mathbf{A}_{_{1-}}\!\equiv\!( & k_{z},k_{x}^{3}\!-\!3k_{x}k_{y}^{2},k_{z}^{3},k_{z}^{2}(k_{x}^{3}\!-\!3k_{x}k_{y}^{2}),k_{z}^{5}),\!\\
\!\!\!\mathbf{A}_{_{2-}}\!\equiv\!( & k_{y}(3k_{x}^{2}-k_{y}^{2}),k_{y}(3k_{x}^{2}-k_{y}^{2})k_{z}^{2}),\!\\
\!\!\!\mathbf{E}_{-}\!\equiv\!( & k_{x},K_{1}k_{z},\!k_{x}k_{z}^{2},\!(k_{x}^{5}\!-\!10k_{x}^{3}k_{y}^{2}+\!5k_{x}k_{y}^{4}),\!\nonumber \\
( & K_{1}^{2}\!-\!4K_{2}^{2})k_{z},K_{1}k_{z}^{3},k_{x}k_{z}^{4}),\!\\
\!\!\!\mathbf{E}_{-}^{\prime}\!\equiv\!( & -k_{y},2K_{3},-k_{y}k_{z}^{2},(k_{y}^{5}\!-\!10k_{x}^{2}k_{y}^{3}\!+\!5k_{x}^{4}k_{y}),\!\nonumber \\
 & -4K_{1}K_{3},2K_{3}k_{z}^{2},-k_{y}k_{z}^{4}\big),\!\label{eq:EprimeNegKK}
\end{align}
where $K_{3}\equiv k_{x}k_{y}k_{z}$. Note that the coefficients of momentum basis in $\mathbf{E}_{\pm}^{\prime}$ are imposed by $A_{1+}$ constraint. Moreover, in Eqs.~(\ref{eq:A1pA1p})--(\ref{eq:EprimeNegKK}),
we have defined a multi-element vector of arbitrary coefficients to
control the strength of momentum dependent polynomials in Eqs.~(\ref{A1pluskk})--(\ref{eq:EprimeNegKK})
as
\begin{align}
\mathbf{P}_{i} & =(P_{1}^{(i)},P_{2}^{(i)},P_{3}^{(i)},P_{4}^{(i)},P_{5}^{(i)}),\\
\mathbf{N}_{i} & =(n_{1}^{(i)},n_{2}^{(i)},n_{3}^{(i)},n_{4}^{(i)},n_{5}^{(i)}),\\
\mathbf{T}_{i} & =(T_{1}^{(i)},T_{2}^{(i)},T_{3}^{(i)},T_{4}^{(i)},T_{5}^{(i)}),\\
\mathbf{R}_{i} & =(R_{1}^{(i)},R_{2}^{(i)}),\\
\mathbf{G}_{i} & =(G_{1}^{(i)},G_{2}^{(i)},G_{3}^{(i)},G_{4}^{(i)},G_{5}^{(i)},G_{6}^{(i)},G_{7}^{(i)}).
\end{align}
Importantly, the above coefficients can be derived through matching
the $\mathbf{k} \cdot \mathbf{p}$ model to the results of  density functional theory.

\section{\label{All_Hamils}Explicit form of effective Hamiltonians}
In this part, we summarize the effective model Hamiltonians in $j\in\{1/2,3/2,5/2\}$ basis. In $j=1/2$ basis, the full Hamiltonian includes four TAM basis matrices  yielding
\selectlanguage{english}%
\begin{align}
\!\!\!\hat{H}_{j=1/2}(\mathbf{k}) & \!=\!C_{0,\mathbf{k}}\hat{J}_{0}\!+\!C_{x,\mathbf{k}}\hat{J}_{x}\!+\!C_{y,\mathbf{k}}\hat{J}_{y}\!+\!C_{z,\mathbf{k}}\hat{J}_{z},\!\!
\end{align}
where the momentum dependent polynomials are 
\begin{align}
C_{0,\mathbf{k}} & =n_{0}+n_{1}(k_{x}^{2}+k_{y}^{2})+n_{2}k_{z}^{2},\\
C_{x,\mathbf{k}} & =-\gamma_{1}k_{y}+\gamma_{2}2k_{x}k_{y}k_{z}-\gamma_{3}k_{y}k_{z}^{2}\nonumber \\
 & +\gamma_{4}(k_{y}^{5}-\!10k_{x}^{2}k_{y}^{3}+5k_{x}^{4}k_{y}\!),\\
C_{y,\mathbf{k}} & =+\gamma_{1}k_{x}+\gamma_{2}(k_{x}^{2}-k_{y}^{2})k_{z}+\gamma_{3}k_{x}k_{z}^{2}\nonumber \\
 & +\gamma_{4}(k_{x}^{5}\!-\!10k_{x}^{3}k_{y}^{2}\!+\!5k_{x}k_{y}^{4}),\\
C_{z,\mathbf{k}} & =m_{1}\big(3k_{y}k_{x}^{2}-k_{y}^{3}\big)+m_{2}\big((3k_{y}k_{x}^{2}\!-\!k_{y}^{3})k_{z}^{2}\big).
\end{align}
Here $n_{0}$ denotes the Fermi energy, $n_{1}$ ($n_{2}$) is the
strength of the in-plane (out-of-plane) kinetic term, $C_{x(y),\mathbf{k}}$
encodes the components of Rashba SOC up to fifth-order momenta, where
$\gamma_{1,2,3,4}$ control the strength of each term, and $C_{z,\mathbf{k}}$
denotes the SOC contribution with out-of-plane spin texture components. \Masoud{In the $j=1/2$ basis, the TAM matrices become
\begin{align}
\!\hat{J}_{x}\! & =\!\frac{1}{2}\begin{pmatrix}0 & 1\\
1 & 0
\end{pmatrix},\!\! & \!\!\hat{J}_{y} & \!=\!\frac{1}{2}\begin{pmatrix}0 & -i\\
i & 0
\end{pmatrix},\!\! & \!\hat{J}_{z}\! & =\!\frac{1}{2}\begin{pmatrix}1 & 0\\
0 & -1
\end{pmatrix}.\!\!
\end{align}
In this basis, the explicit matrix form of the Hamiltonian is given by
\begin{widetext}

\begin{align}
\hat{H}_{j=1/2}(\mathbf{k})\! & =\!(\mu\!+\!n_{1}|\mathbf{k}_{\perp}|^{2}\!+\!n_{2}k_{z}^{2})\left(\begin{array}{cc}
1 & 0\\
0 & 1
\end{array}\right)\nonumber \\
 & +\left(\begin{array}{cc}
(m_{1}+m_{2}k_{z}^{2})\text{Im}(k_{+}^{3}) & -i(\gamma_{1}k_{-}+\gamma_{2}k_{+}^{2}k_{z}+\gamma_{3}k_{-}k_{z}^{2}+\gamma_{4}k_{+}^{5})\\
i(\gamma_{1}k_{+}+\gamma_{2}k_{-}^{2}k_{z}+\gamma_{3}k_{+}k_{z}^{2}+\gamma_{4}k_{-}^{5}) & -(m_{1}+m_{2}k_{z}^{2})\text{Im}(k_{+}^{3})
\end{array}\right),
\end{align}
\end{widetext}
where the off-diagonal terms correspond to spin--orbit coupling contributions
ranging from first to fifth order, with the first-order term representing
the conventional Rashba SOC, while the diagonal terms describe hexagonal warping. Here and throughout, we use the shorthand notation \(k_{\pm}=k_x\pm i k_y\).}

Moreover, the effective Hamiltonian admits a much richer operator
structure for higher total angular momentum. In a $(2j+1)$-dimensional Hilbert space, the number of linearly independent Hermitian matrices is $(2j+1)^{2}$. Hence, there are $16$ independent operator channels for $j=3/2$ and $36$ for $j=5/2$. Accordingly, we expand the Hamiltonian as
\begin{align}
\hat{H}(\mathbf{k}) & =\mathcal{A}_{1,\mathbf{k}}\hat{J}_{0}+\mathcal{A}_{2,\mathbf{k}}\hat{J}_{z}^{2}\nonumber \\
 & +\varLambda_{1,\mathbf{k}}\hat{J}_{y}+\Upsilon_{1,\mathbf{k}}\hat{J}_{x}\nonumber \\
 & +\varLambda_{2,\mathbf{k}}\lceil\hat{J}_{y}\hat{J}_{z}^{2}\rfloor+\Upsilon_{2,\mathbf{k}}\lceil\hat{J}_{x}\hat{J}_{z}^{2}\rfloor\nonumber \\
 & +\varLambda_{3,\mathbf{k}}\lceil\hat{J}_{x}\hat{J}_{y}\hat{J}_{z}\rfloor+\frac{1}{2}\Upsilon_{3,\mathbf{k}}\llbracket(\hat{J}_{y}^{2}-\hat{J}_{x}^{2})\hat{J}_{z}\rrbracket\nonumber \\
 & +\mathcal{B}_{1,\mathbf{k}}\hat{J}_{z}+\zeta_{1}k_{z}(3\llbracket\hat{J}_{y}\hat{J}_{x}^{2}\rrbracket-\hat{J}_{y}^{3})\nonumber \\
 & +\mathcal{B}_{2,\mathbf{k}}\hat{J}_{z}^{3}+\mathcal{B}_{3,\mathbf{k}}(\hat{J}_{x}^{3}-3\llbracket\hat{J}_{x}\hat{J}_{y}^{2}\rrbracket)\nonumber \\
 & +\zeta_{2}k_{z}\lceil(3\llbracket\hat{J}_{y}\hat{J}_{x}^{2}\rrbracket-\hat{J}_{y}^{3})\hat{J}_{z}^{2}\rfloor+\cdots,
 \label{Full_Ham_App}
\end{align}
where 
\begin{equation}
\mathcal{B}_{i,\mathbf{k}}\!=\!m_{1,i}k_{y}(3k_{x}^{2}\!-\!k_{y}^{2})\!+\!m_{2,i}k_{y}(3k_{x}^{2}\!-\!k_{y}^{2})k_{z}^{2},
\end{equation}
and $\mathcal{A}_{i,\mathbf{k}}$, $\varLambda_{i,\mathbf{k}}$ and
$\Upsilon_{i,\mathbf{k}}$ are defined in Eqs.~\eqref{Ai_coef}, \eqref{Lamdai}, and \eqref{Upsiloni},
respectively. All remaining symmetry-allowed contributions are obtained
systematically by combining momentum-basis polynomials with TAM-tensor
operators such that the resulting terms transform according to the
$A_{1,+}$ irrep.

\Masoud{To obtain the explicit form of the Hamiltonian in the $j=3/2$ basis, we substitute the following TAM matrices into Eq.~\eqref{Full_Ham_App}
\begin{align}
J_{x} & =\begin{pmatrix}0 & \frac{\sqrt{3}}{2} & 0 & 0\\
\frac{\sqrt{3}}{2} & 0 & 1 & 0\\
0 & 1 & 0 & \frac{\sqrt{3}}{2}\\
0 & 0 & \frac{\sqrt{3}}{2} & 0
\end{pmatrix},\\
J_{y} & =\begin{pmatrix}0 & -i\frac{\sqrt{3}}{2} & 0 & 0\\
i\frac{\sqrt{3}}{2} & 0 & -i & 0\\
0 & i & 0 & -i\frac{\sqrt{3}}{2}\\
0 & 0 & i\frac{\sqrt{3}}{2} & 0
\end{pmatrix},\\
J_{z} & =\frac{1}{2}\text{diag}(3,1,-1,-3).
\end{align}
Then, expanding Eq.~\eqref{Full_Ham_App} in powers of momenta yields
\begin{align}
\hat{H}(\mathbf{k})\! & =\!(\mu_{1}\!+\!\alpha_{1,1}|\mathbf{k}_{\perp}|^{2}\!+\!\alpha_{1,2}k_{z}^{2})\left(\!\begin{array}{cccc}
1 & 0 & 0 & 0\\
0 & 1 & 0 & 0\\
0 & 0 & 1 & 0\\
0 & 0 & 0 & 1
\end{array}\!\right)\!\!\nonumber \\
 & +\!(\mu_{2}\!+\!\alpha_{2,1}|\mathbf{k}_{\perp}|^{2}\!+\!\alpha_{2,2}k_{z}^{2})\left(\!\begin{array}{cccc}
\frac{9}{4} & 0 & 0 & 0\\
0 & \frac{1}{4} & 0 & 0\\
0 & 0 & \frac{1}{4} & 0\\
0 & 0 & 0 & \frac{9}{4}
\end{array}\!\right)\!\!\nonumber \\
 & +\!\sum_{\nu=1}\!\hat{H}_{\text{SOC}}^{(2\nu-1)}(\mathbf{k}),\!\!\label{HamiLOrder}
\end{align}
where $\mu_{1}$ and $\mu_{2}$ are the zeroth-order coefficients
corresponding to the chemical potential and the energy shift induced
by even-parity SOC, respectively, $\alpha_{1,1}$ and $\alpha_{1,2}$
describe the in-plane and out-of-plane kinetic-energy contributions,
while $\alpha_{2,1}$ and $\alpha_{2,2}$ denote their TAM-dependent
counterparts arising from the even-parity SOC term, and $\hat{H}_{\mathrm{SOC}}^{(2\nu-1)}(\mathbf{k})$
represents the odd-parity SOC Hamiltonian of order $2\nu-1$; the
first-order SOC terms are given by  
\begin{align}
\hat{H}_{\text{SOC}}^{(1)}(\mathbf{k})\! & =\!\gamma_{1,1}\left(\begin{array}{cccc}
0 & -i\frac{\sqrt{3}}{2}k_{-} & 0 & 0\\
i\frac{\sqrt{3}}{2}k_{+} & 0 & -ik_{-} & 0\\
0 & ik_{+} & 0 & -i\frac{\sqrt{3}}{2}k_{-}\\
0 & 0 & i\frac{\sqrt{3}}{2}k_{+} & 0
\end{array}\right)\!\nonumber \\
 & +\!\gamma_{1,2}\left(\begin{array}{cccc}
0 & -i\frac{5\sqrt{3}}{8}k_{-} & 0 & 0\\
i\frac{5\sqrt{3}}{8}k_{+} & 0 & -i\frac{1}{4}k_{-} & 0\\
0 & i\frac{1}{4}k_{+} & 0 & -i\frac{5\sqrt{3}}{8}k_{-}\\
0 & 0 & i\frac{5\sqrt{3}}{8}k_{+} & 0
\end{array}\right)\nonumber \\
 & +\frac{\sqrt{3}}{4}\gamma_{1,3}\left(\begin{array}{cccc}
0 & 0 & -ik_{+} & 0\\
0 & 0 & 0 & ik_{+}\\
ik_{-} & 0 & 0 & 0\\
0 & -ik_{-} & 0 & 0
\end{array}\right)\!\nonumber \\
 & +\!\zeta_{1}\left(\begin{array}{cccc}
0 & 0 & 0 & -3ik_{z}\\
0 & 0 & 0 & 0\\
0 & 0 & 0 & 0\\
3ik_{z} & 0 & 0 & 0
\end{array}\right),\!
\end{align}
where the first term corresponds to the conventional Rashba SOC, while
the remaining channels arise from nontrivial spin-tensor matrices
in the $j=3/2$ basis. In addition, the third-order SOC terms are given by 
\begin{align}
\hat{H}_{\text{SOC}}^{(3)}(\mathbf{k}) & =\gamma_{2,1}k_{z}\left(\begin{array}{cccc}
0 & -i\frac{\sqrt{3}}{2}k_{+}^{2} & 0 & 0\\
i\frac{\sqrt{3}}{2}k_{-}^{2} & 0 & -ik_{+}^{2} & 0\\
0 & ik_{-}^{2} & 0 & -i\frac{\sqrt{3}}{2}k_{+}^{2}\\
0 & 0 & i\frac{\sqrt{3}}{2}k_{-}^{2} & 0
\end{array}\right)\nonumber \\
 & +\gamma_{3,1}k_{z}^{2}\left(\begin{array}{cccc}
0 & -i\frac{\sqrt{3}}{2}k_{-} & 0 & 0\\
i\frac{\sqrt{3}}{2}k_{+} & 0 & -ik_{-} & 0\\
0 & ik_{+} & 0 & -i\frac{\sqrt{3}}{2}k_{-}\\
0 & 0 & i\frac{\sqrt{3}}{2}k_{+} & 0
\end{array}\right)\nonumber \\
 & +\gamma_{2,2}k_{z}\left(\begin{array}{cccc}
0 & -i\frac{5\sqrt{3}}{8}k_{+}^{2} & 0 & 0\\
i\frac{5\sqrt{3}}{8}k_{-}^{2} & 0 & -\frac{i}{4}k_{+}^{2} & 0\\
0 & \frac{i}{4}k_{-}^{2} & 0 & -i\frac{5\sqrt{3}}{8}k_{+}^{2}\\
0 & 0 & i\frac{5\sqrt{3}}{8}k_{-}^{2} & 0
\end{array}\right)\nonumber \\
 & +\gamma_{3,2}k_{z}^{2}\left(\begin{array}{cccc}
0 & -i\frac{5\sqrt{3}}{8}k_{-} & 0 & 0\\
i\frac{5\sqrt{3}}{8}k_{+} & 0 & -\frac{i}{4}k_{-} & 0\\
0 & \frac{i}{4}k_{+} & 0 & -i\frac{5\sqrt{3}}{8}k_{-}\\
0 & 0 & i\frac{5\sqrt{3}}{8}k_{+} & 0
\end{array}\right)\nonumber \\
 & +\frac{\sqrt{3}}{4}\gamma_{2,3}k_{z}\left(\begin{array}{cccc}
0 & 0 & -ik_{-}^{2} & 0\\
0 & 0 & 0 & ik_{-}^{2}\\
ik_{+}^{2} & 0 & 0 & 0\\
0 & -ik_{+}^{2} & 0 & 0
\end{array}\right)\nonumber \\
 & +\frac{\sqrt{3}}{4}\gamma_{3,3}k_{z}^{2}\left(\begin{array}{cccc}
0 & 0 & -ik_{+} & 0\\
0 & 0 & 0 & ik_{+}\\
ik_{-} & 0 & 0 & 0\\
0 & -ik_{-} & 0 & 0
\end{array}\right)\nonumber \\
 & +m_{1,1}\left(\begin{array}{cccc}
\frac{3}{2} & 0 & 0 & 0\\
0 & \frac{1}{2} & 0 & 0\\
0 & 0 & -\frac{1}{2} & 0\\
0 & 0 & 0 & \frac{3}{2}
\end{array}\right)k_{y}(3k_{x}^{2}-k_{y}^{2})\nonumber \\
 & +m_{1,2}\left(\begin{array}{cccc}
\frac{27}{8} & 0 & 0 & 0\\
0 & \frac{1}{8} & 0 & 0\\
0 & 0 & -\frac{1}{8} & 0\\
0 & 0 & 0 & \frac{27}{8}
\end{array}\right)k_{y}(3k_{x}^{2}-k_{y}^{2})\nonumber \\
 & +3m_{1,3}\left(\begin{array}{cccc}
0 & 0 & 0 & 1\\
0 & 0 & 0 & 0\\
0 & 0 & 0 & 0\\
1 & 0 & 0 & 0
\end{array}\right)k_{y}(3k_{x}^{2}-k_{y}^{2}).
\end{align}

Finally, the fifth-order terms are given by
\begin{align}
\!\hat{H}_{\text{SOC}}^{(5)}(\mathbf{k}) & \!=\!\gamma_{4,1}\left(\begin{array}{cccc}
0 & -i\frac{\sqrt{3}}{2}k_{+}^{5} & 0 & 0\\
i\frac{\sqrt{3}}{2}k_{-}^{5} & 0 & -ik_{+}^{5} & 0\\
0 & ik_{-}^{5} & 0 & -i\frac{\sqrt{3}}{2}k_{+}^{5}\\
0 & 0 & i\frac{\sqrt{3}}{2}k_{-}^{5} & 0
\end{array}\right)\!\nonumber \\
 & +\!\gamma_{4,2}\left(\begin{array}{cccc}
0 & -i\frac{5\sqrt{3}}{8}k_{+}^{5} & 0 & 0\\
i\frac{5\sqrt{3}}{8}k_{-}^{5} & 0 & -\frac{i}{4}k_{+}^{5} & 0\\
0 & \frac{i}{4}k_{-}^{5} & 0 & -i\frac{5\sqrt{3}}{8}k_{+}^{5}\\
0 & 0 & i\frac{5\sqrt{3}}{8}k_{-}^{5} & 0
\end{array}\right)\nonumber \\
 & +\frac{\sqrt{3}}{4}\gamma_{4,3}\left(\begin{array}{cccc}
0 & 0 & -ik_{-}^{5} & 0\\
0 & 0 & 0 & ik_{-}^{5}\\
ik_{+}^{5} & 0 & 0 & 0\\
0 & -ik_{+}^{5} & 0 & 0
\end{array}\right)\!\nonumber \\
 & +m_{2,1}\left(\begin{array}{cccc}
\frac{3}{2} & 0 & 0 & 0\\
0 & \frac{1}{2} & 0 & 0\\
0 & 0 & -\frac{1}{2} & 0\\
0 & 0 & 0 & \frac{3}{2}
\end{array}\right)\!k_{y}(3k_{x}^{2}-k_{y}^{2})k_{z}^{2}\!\!\nonumber \\
 & +\!m_{2,2}\left(\begin{array}{cccc}
\frac{27}{8} & 0 & 0 & 0\\
0 & \frac{1}{8} & 0 & 0\\
0 & 0 & -\frac{1}{8} & 0\\
0 & 0 & 0 & \frac{27}{8}
\end{array}\right)\!k_{y}(3k_{x}^{2}-k_{y}^{2})k_{z}^{2}\!\!\nonumber \\
 & +3m_{2,3}\left(\begin{array}{cccc}
0 & 0 & 0 & 1\\
0 & 0 & 0 & 0\\
0 & 0 & 0 & 0\\
1 & 0 & 0 & 0
\end{array}\right)k_{y}(3k_{x}^{2}-k_{y}^{2})k_{z}^{2}.
\end{align}
The material-dependent coefficients in Eq.~($\ref{HamiLOrder}$)
can be determined by fitting the model to density functional theory
simulations of the band structure. Note that, similarly to the $j=3/2$ basis, the expansion in Eq.~\eqref{HamiLOrder} can also be carried out in the $j=5/2$ basis. In this case, one has to use the corresponding $6\times6$ matrices given in Eqs.~\eqref{Jz52} and \eqref{Jx52}--\eqref{Jy52}.}

\section{\label{FPCs}First-principle calculations}
In this section, we briefly explain how to obtain the TAM texture from first-principle calculations. In practice, the Bloch spinors $|u_{n\mathbf{k}}\rangle$ can be obtained
from a noncollinear SOC calculation in VASP \cite{VASP_1996,VASP_1999}. 
 For noncollinear SOC case, the three components of the projected spin magnetization (Pauli-matrix expectation values) directly yield the usual $j=1/2$ spin texture \cite{vaspwiki_procar_2025}. To obtain TAM textures in
a target $j$-manifold (e.g. $j=3/2$ for $p$-orbital), one should reconstruct the local spinor coefficients in the orbital-spin product basis $w_{im}^{n\mathbf{k}}=\langle i,lm_{l}\otimes m_{s}|u_{n\mathbf{k}}\rangle$
, and rotate to the local $|jm_{j}\rangle$ basis using Clebsch--Gordan coefficients (CGCs).
Finally, we can evaluate the projected expectation value $\mathbf{J}$.

The numerical simulation results in discrete $\mathbf{k}$-points,
energy bands and how many ions are taken into account. The Wannier
basis includes orbital and spin-$1/2$ degrees of freedom. Thus, we
need to obtain the projection of Wannier basis onto the TAM basis.
The TAM operator is $\hat{\mathbf{J}}=\hat{\mathbf{L}}+\hat{\mathbf{S}}$,
and for any normalized state $\langle\hat{\mathbf{J}}\rangle_{n\mathbf{k}}=\langle u_{n\mathbf{k}}|\hat{\mathbf{L}}|u_{n\mathbf{k}}\rangle+\langle u_{n\mathbf{k}}|\hat{\mathbf{S}}|u_{n\mathbf{k}}\rangle$
where $|u_{n\mathbf{k}}\rangle$ is the Bloch spinor wave function.
A $j>1/2$ manifold requires a projected TAM texture $\langle\hat{\mathbf{J}}\rangle_{n\mathbf{k}}^{(j)}=\langle u_{n\mathbf{k}}|P_{j}\,\hat{\mathbf{J}}\,P_{j}|u_{n\mathbf{k}}\rangle/\langle u_{n\mathbf{k}}|P_{j}|u_{n\mathbf{k}}\rangle$.
 The component
of conventional spin-$1/2$ texture can be obtained by $\langle\hat{S}_{i}\rangle_{n\mathbf{k}}=(1/2)\sum_{\sigma\sigma^{\prime}}(\hat{\sigma}_{i})_{\sigma\sigma^{\prime}}\langle u_{n\mathbf{k}}^{\sigma}|\hat{P}|u_{n\mathbf{k}}^{\sigma^{\prime}}\rangle$
where $|u_{n\mathbf{k}}^{\sigma}\rangle$ is the weight for spinor
component with spin degree of freedom $\sigma\in\{\uparrow,\downarrow\}$,
$\hat{\sigma}_{i}$ is the $i$th Pauli matrix where $i\in\{x,y,z\}$,
and $\hat{P}$ denotes the local/orbital projector. For a heavy-element
$p$ or $d$ manifold, the expansion of Bloch spinor in a localized
basis takes the form $|lm_{l}\otimes\sigma\rangle\equiv|lm_{l}\rangle\otimes|\sigma\rangle$
in density functional theory (DFT). In this case, the periodic part
of the Bloch state is
\begin{equation}
|u_{n\mathbf{k}}\rangle=\sum_{m_{l},\sigma}c_{lm_{l}}^{sm_{s}}|lm_{l}\otimes sm_{s}\rangle,
\end{equation}
where $c_{lm_{l}}^{sm_{s}}=\langle lm_{l}\otimes sm_{s}|u_{n\mathbf{k}}\rangle$
denotes the local orbital-spin weight for the projected Bloch state. The projection of spin-$1/2$ texture vector can be obtained by summation of its projection onto the real space basis as $\langle\mathbf{S}\rangle_{n\mathbf{k}}=(\hbar/2)\sum_{i}\langle u_{n\mathbf{k}}|\hat{P}_{i}\hat{\mathbf{\sigma}}\hat{P}_{i}|u_{n\mathbf{k}}\rangle$
 where $\hat{P}_{i}$ is the matrix of projection onto the $i$th site.

Analogously, the TAM texture vector can be obtained likewise. For
$p$-orbital $(l=1)$ with strong SOC, the relevant high-$j$ manifold
is the quartet $j=3/2$ TAM. In this case, the relevant TAM texture
is the average value of $\hat{\mathbf{J}}=(\hat{J}_{x},\hat{J}_{y},\hat{J}_{z})$
projected to the $j=3/2$ subspace at site $i$ as defined by $\langle\hat{\mathbf{J}}\rangle_{ijn\mathbf{k}}$.
This projection is $\hat{P}_{ij}=\sum_{m_{j}}|ijm_{j}\rangle\langle ijm_{j}|$
where $\hat{P}_{ij}$ is Hermitian $\hat{P}_{ij}^{\dagger}=\hat{P}_{ij}$
and fulfills $\hat{P}_{ij}^{2}=\hat{P}_{ij}$. This projector keeps
only the $j$ degrees of freedom on the site index $i$. Under this
projection the normalized Bloch state takes the form $|\psi_{ij}^{n\mathbf{k}}\rangle=(\hat{P}_{ij}/\langle\hat{P}_{ij}\rangle_{n\mathbf{k}}^{1/2})|u_{n\mathbf{k}}\rangle$
where the site-resolved TAM texture becomes 
\begin{align}
\langle\hat{\mathbf{J}}\rangle_{n\mathbf{k}}^{ij} & \equiv\frac{1}{\langle\hat{P}_{ij}\rangle_{n\mathbf{k}}}\langle u_{n\mathbf{k}}|\hat{P}_{ij}\hat{\mathbf{J}}\hat{P}_{ij}|u_{n\mathbf{k}}\rangle.\label{J_firstprinciple}
\end{align}
In the DFT/Wannier basis the local-$j$ state can be obtained through
the expansion $|ijm_{j}\rangle=\sum_{m_{l}m_{s}}C_{lm_{l}m_{s}}^{jm_{j}}|ilm_{l}\otimes m_{s}\rangle$
where $C_{lm_{l}m_{s}}^{jm_{j}}=\langle lm_{l}m_{s}|jm_{j}\rangle$
is the CGC and $|ilm_{l}\otimes m_{s}\rangle$
is the Wannier state labeled with site $i$, orbital $l$ ($m_{l}$
quantum number), and spin $m_{s}\in\{\uparrow,\downarrow\}$ degrees
of freedom . Under this linear expansion, the local $j-$space projector
is equivalent to 
\begin{equation}
\hat{P}_{ij}=\sum_{mm^{\prime}}\mathfrak{C}_{m}^{m^{\prime}}|im\rangle\langle im^{\prime}|,
\end{equation}
where $m=(m_{l},m_{s})$ and $|im\rangle\equiv|ilm_{l}\otimes m_{s}\rangle$,
$\mathcal{C}_{mm^{\prime}}^{m_{j}}\equiv C_{m}^{m_{j}}(C_{m^{\prime}}^{m_{j}})^{*}$
is the product of CGCs, and $\mathfrak{C}_{m}^{m^{\prime}}=\sum_{m_{j}}\mathcal{C}_{mm^{\prime}}^{m_{j}}$.
In the case, the site-resolved average of TAM becomes
\begin{align}
\!\!\!\!\langle\hat{\mathbf{J}}\rangle_{n\mathbf{k}}^{ij} & \!=\!\frac{1}{\mathcal{N}}\!\!\sum_{m_{1}m_{2}}\!\sum_{m_{3}m_{4}}\!\mathfrak{C}_{m_{1}}^{m_{2}}\mathfrak{C}_{m_{3}}^{m_{4}}(w_{im_{1}}^{n\mathbf{k}})^{*}w_{im_{4}}^{n\mathbf{k}}(J_{i\alpha})_{m_{2}m_{3}},\!\!\!
\end{align}
where $\mathcal{N}$ is the normalization factor, $w_{im}^{n\mathbf{k}}=\langle im|u_{n\mathbf{k}}\rangle$, and
the matrix element for the TAM matrices is 
\begin{equation}
\left(J_{i\alpha}\right)_{mm^{\prime}}=\left(L_{i\alpha}\right)_{m_{l}m_{l}^{\prime}}\delta_{m_{s}m_{s}^{\prime}}+\delta_{m_{l}m_{l}^{\prime}}\left(s_{i\alpha}\right)_{m_{s}m_{s}^{\prime}},
\end{equation}
where $\left(O_{i\alpha}\right)_{mm^{\prime}}$ with $O\in\{L,s\}$
denotes the matrix element of local orbital and spin.

\bibliography{Final_biblio_multipolar}
\end{document}